\documentclass[article]{imsart}

\RequirePackage[OT1]{fontenc}
\RequirePackage{amsthm,amsmath,enumerate,mathtools,verbatim}
\RequirePackage{natbib}
\RequirePackage[colorlinks,citecolor=blue,urlcolor=blue]{hyperref}
\usepackage{tabu}

\arxiv{arXiv:0000.0000}

\startlocaldefs
\numberwithin{equation}{section}
\theoremstyle{plain}

\usepackage{graphicx}
\usepackage{enumerate}

\usepackage{amsmath,amssymb,latexsym}
\usepackage{natbib}
\usepackage{times}
\usepackage{bm}
\usepackage{mathrsfs}           

\usepackage{verbatim}
\usepackage{changebar}
\usepackage{stfloats}	
\usepackage{color}
\usepackage{xcolor}
\usepackage{multirow}

\usepackage[utf8]{inputenc}
\usepackage{fourier}
\usepackage{array}
\usepackage{makecell}

\def\Z{{\mathbb Z}}

\def\S1{{\mathscr S}}
\def\M1{{\mathscr M}}

\newcommand{\beginsupplement}{%
        \setcounter{table}{0}
        \renewcommand{\thetable}{S\arabic{table}}%
        \setcounter{figure}{0}
        \renewcommand{\thefigure}{S\arabic{figure}}%
     }

\def\beq{\begin{equation}}
\def\eeq{\end{equation}}
\def\beqa{\begin{eqnarray}}
\def\eeqa{\end{eqnarray}}
\def\beqann{\begin{eqnarray*}}
\def\eeqann{\end{eqnarray*}}
\def\bcb{\begin{changebar}}
\def\ecb{\end{changebar}}

\endlocaldefs

\begin{document}

\begin{frontmatter}
\title{Wavelet spectral testing: application to nonstationary circadian rhythms}
\runtitle{Wavelet spectral testing for nonstationary circadian rhythms}

\begin{aug}
\author{\fnms{Jessica K. } \snm{Hargreaves}},
\author{\fnms{Marina I. } \snm{Knight}},
\author{\fnms{Jon W. } \snm{Pitchford}},
\author{\fnms{Rachael J.} \snm{Oakenfull}},
\author{\fnms{Sangeeta } \snm{ Chawla}},
\author{\fnms{Jack} \snm{Munns}}
\and
\author{\fnms{Seth J. } \snm{Davis}}

\runauthor{J. K. Hargreaves et al.}

\end{aug}

\begin{abstract}
Rhythmic data are ubiquitous in the life sciences. Biologists need reliable statistical tests to identify whether a particular experimental treatment has caused a significant change in a rhythmic signal. When these signals display nonstationary behaviour, as is common in many biological systems, the established methodologies may be misleading. Therefore, there is a real need for new methodology that enables the formal comparison of nonstationary processes. As circadian behaviour is best understood in the spectral domain, here we develop novel hypothesis testing procedures in the (wavelet) spectral domain, embedding replicate information when available. The data are modelled as realisations of locally stationary wavelet processes, allowing us to define and rigorously estimate their evolutionary wavelet spectra. Motivated by three complementary applications in circadian biology, our new methodology allows the identification of three specific types of spectral difference. We demonstrate the advantages of our methodology over alternative approaches, by means of a comprehensive simulation study and real data applications, using both published and newly generated circadian datasets. In contrast to the current standard methodologies, our method successfully identifies differences within the motivating circadian datasets, and facilitates wider ranging analyses of rhythmic biological data in general.
\end{abstract}

\begin{keyword}[class=MSC]
\kwd[Primary ]{62M10}
\kwd{60G18}
\kwd[; secondary ]{60-08}
\end{keyword}

\begin{keyword}
\kwd{wavelets}
\kwd{spectral decomposition}
\kwd{hypothesis testing}
\kwd{circadian rhythms}
\end{keyword}

\begin{keyword}[class=Funding]
\kwd[\textbf{Affiliation:} Departments of Mathematics and Biology, University of York]{}
\end{keyword}

\begin{keyword}[class=Funding]
\kwd[\textbf{Funding:} This work was supported by EPSRC. Circadian work in the SJD group is currently funded by BBSRC awards BB/M000435/1 and BB/N018540/1]{}
\end{keyword}

\end{frontmatter}

\section{Introduction}\label{sec:intro}

Almost all species exhibit changes in their behaviour between day and night \citep{bell2005circadian}. These daily rhythms (known as `circadian rhythms') are the result of an internal timekeeping system, in response to daily changes in the physical environment \citep{vitaterna2001overview, minors2013circadian}. The `circadian clock' enhances survival by directing anticipatory changes in physiology synchronised with environmental fluctuations. When an organism is deprived of external time cues, its circadian rhythms typically persist qualitatively but may change in detail; the study of these changes can reveal the biochemical reactions underpinning the circadian clock and, at a larger scale, can provide valuable insight into the possible consequences of environmental and ecological challenges \citep{mcclung2006plant, bujdoso2013mathematical}.

\subsection{Motivation}
\label{subsec:Motivation}

In many scientific applications, available data consist of signals with known group memberships and scientists are interested in establishing whether these groups display statistically different behaviour. Our work is motivated by a general problem: biologists need reliable statistical tests to identify whether a particular experimental treatment has caused a significant change in the circadian rhythm. If the changes are limited to period and/or phase then existing Fourier-based theory may be adequate. However, when the changes to the circadian clock are less straightforward, for example involving non-stationarity or changes at multiple scales \citep{hargreaves17:cluster}, the application of these established methods may be conducive to misleading conclusions. The potential value of our approach is illustrated by three complementary examples, encompassing the effect of various salt stresses on plants, the identification of mutations inducing rapid rhythms, and the response of nematode clocks to pharmacological treatment, as described in the following sections. The biological experimental details for each dataset appear in Appendix \ref{App:Expermiental Methods}.

\subsubsection{Lead nitrate dataset (Davis Lab, Biology, University of York)}
\label{subsubsec:Lead Intro}
This dataset (henceforth referred to as the `Lead dataset') is from a broad investigation of whether plant circadian clocks are affected by industrial and agricultural pollutants \citep{foley2005global, senesil1999trace, hargreaves17:cluster, nicholson2003inventory}. Specifically, this experiment asks whether lead affects the \textit{Arabidopsis thaliana} circadian clock and, if so, when and how? Figure \ref{fig:TSLeadpdf} displays the luminescence profiles for both untreated \textit{A. thaliana} plants, as well as for those exposed to lead nitrate.

\subsubsection{Ultradian dataset (Millar Lab, Biology, University of Edinburgh)}
\label{subsubsec:AM Intro}
In order to understand the clock mechanism, a common approach is to mutate a gene and examine the resulting behaviour in response to a variety of stimuli. Figure \ref{fig:TSAMpdf} depicts the luminescence profiles recording plant response to light, for both the control and genetically mutated \textit{A. thaliana} plants \citep{millar2015changing}. Researchers are interested in establishing whether a specific genetic mutation induced high-frequency behaviour (known as `ultradian rhythms') in the laboratory model plant \textit{A. thaliana}.

\subsubsection{Nematode dataset (Chawla Lab, Biology, University of York)}
\label{subsubsec:Worm Intro}
\begin{figure}
\centering
\includegraphics[width=\linewidth]{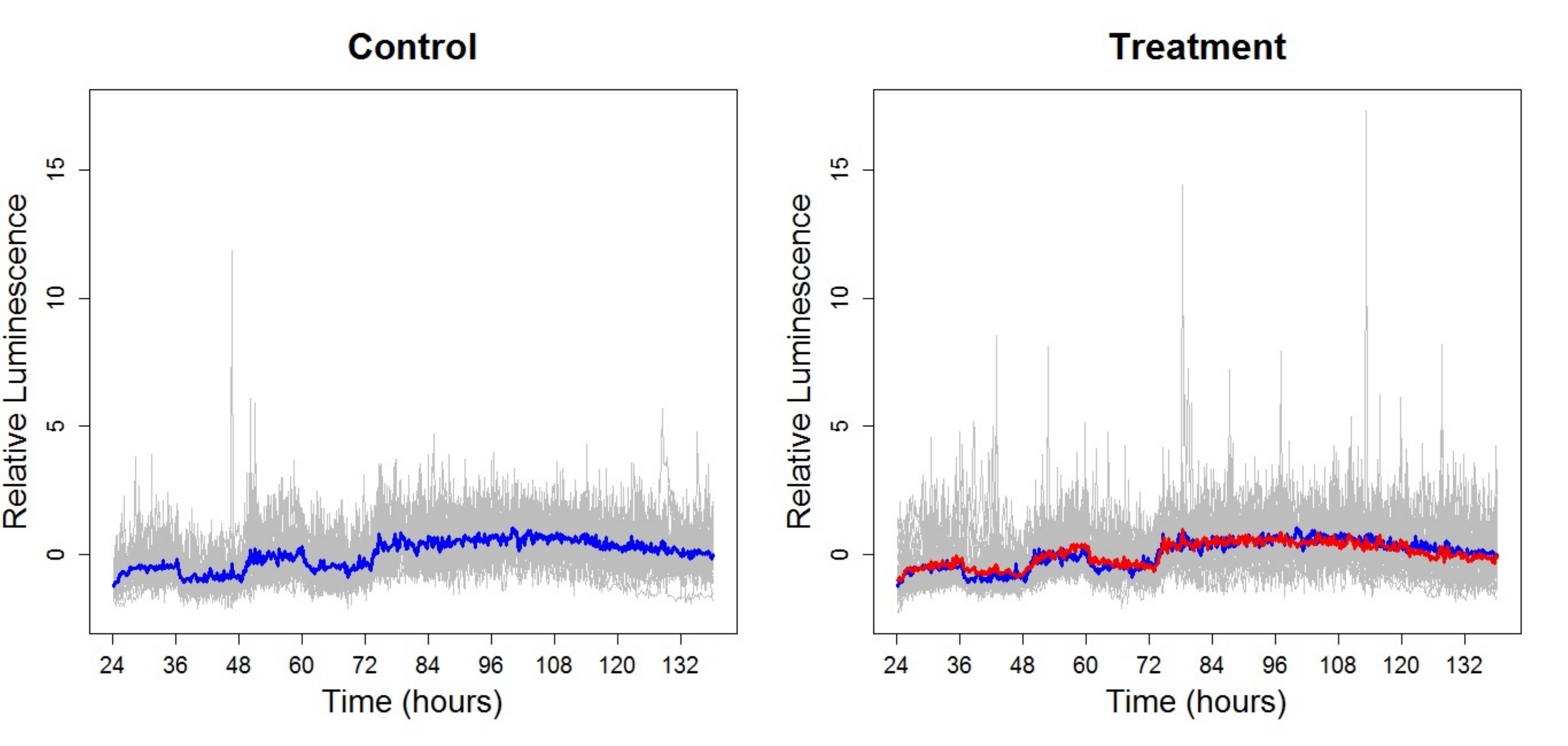}
\caption{\textbf{Nematode dataset:} Luminescence profiles over time for untreated \textit{C. elegans} (Control) and those subjected to a pharmacological treatment (Treatment). Left: Individuals in the control group (in grey) along with the group average (blue). Right: Individuals in the treatment group (in grey) along with the treatment group average (red) and the control group average (blue). Each time series has been standardised to have mean zero.}
\label{fig:TSWormspdf}
\end{figure}

The free-living nematode \textit{Caenorhabditis elegans} is an animal widely used in neuroscience and genetics, but its circadian clock is still poorly understood. To increase understanding of the nematode clock, and potentially uncover rhythmicity not detected by conventional approaches, researchers applied a pharmacological treatment to \textit{C. elegans}, based on evidence that it causes aberrant circadian rhythms in other established mammalian and insect circadian models \citep{kon2015cell, dusik2014map}. Figure \ref{fig:TSWormspdf} depicts the luminescence profiles for both untreated and treated \textit{C. elegans}.

\begin{figure}
\centering
\includegraphics[width=\linewidth]{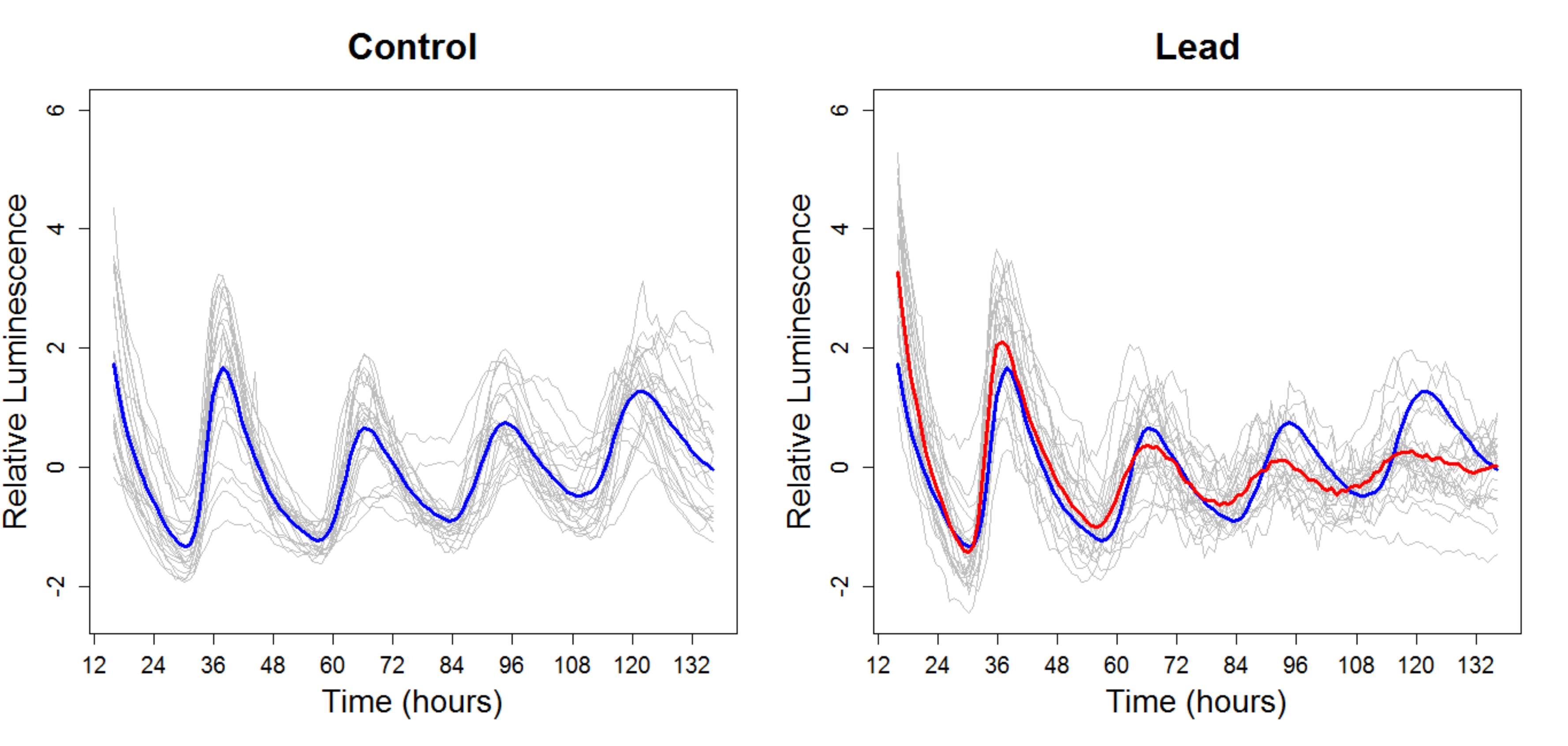}
\caption{\textbf{Lead dataset:} Luminescence profiles over time for untreated \textit{A. thaliana} plants (Control) and those exposed to lead nitrate (Lead). Left: Individuals in the control group (in grey) along with the group average (blue). Right: Individuals in the lead treatment group (in grey) along with the treatment group average (red) and the control group average (blue). Each time series has been standardised to have mean zero.}
\label{fig:TSLeadpdf}
\end{figure}

\begin{figure}
\centering
\includegraphics[width=\linewidth]{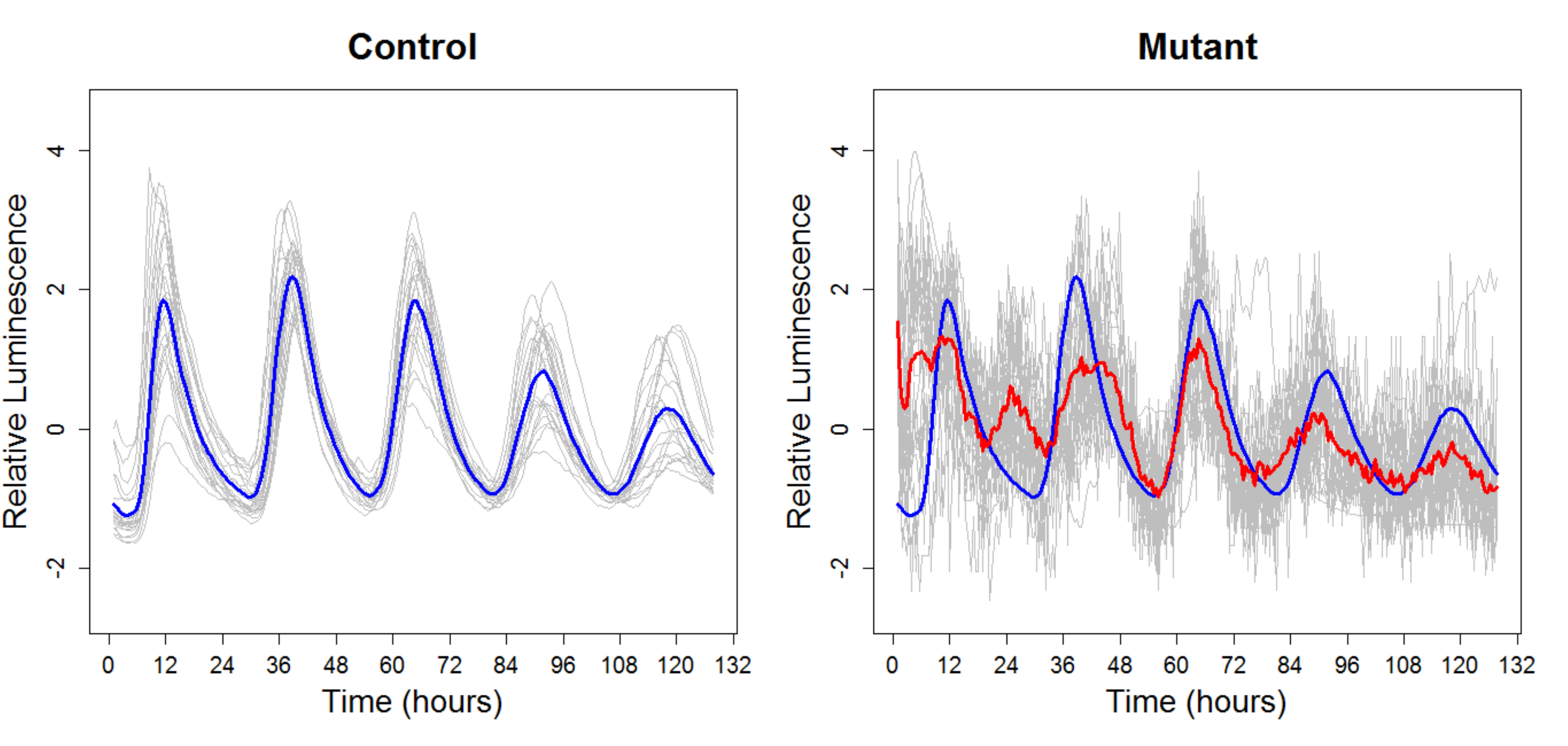}
\caption{\textbf{Ultradian dataset:} Luminescence profiles over time for control and mutant \textit{A. thaliana} plants. Left: Individuals in the control group (in grey) along with the group average (blue). Right: Individuals in the mutant group (in grey) along with the mutant group average (red) and the control group average (blue). Each time series has been standardised to have mean zero.}
\label{fig:TSAMpdf}
\end{figure}

On examining Figures \ref{fig:TSLeadpdf} and \ref{fig:TSAMpdf}, it is visually clear that changes in period and amplitude between the control and test groups occur in both datasets. Figure \ref{fig:TSWormspdf} reveals apparently similar luminescence profiles for both untreated and treated \textit{C. elegans}. Nevertheless, in each experiment less easily quantified or subtle differences between these groups may also exist.

\subsection{Aims and structure of the paper}

Period estimation is central to the analysis of circadian data, with the current standard achieving this using Fourier analysis \citep{zielinski2014strengths, costa2011estimating} via software packages, such as BRASS (Biological Rhythm Analysis Software System \citep{edwards2010quantitative}) or BioDare \citep{moore2014online}. The practitioner estimates the period of the control and treatment groups respectively, and then tests for statistically significant differences (see for example \cite{perea2015modulation}, \cite{costa2011estimating}). Crucially, in all of our motivating examples, such established Fourier-based tests found no significant difference between the groups (see Table \ref{tab:BRASS} in Appendix \ref{App:RDASupMat}).

One obvious limitation of this analysis is that the employed methodology does not typically evaluate the crucial underpinning assumption of data stationarity. In the context examined here, assuming stationarity can be inappropriate \citep{hargreaves17:cluster, leise2013wavelet}, a feature shared by many biological systems \citep{zielinski2014strengths}. For our motivating example datasets, we investigated whether the individual time series are (second-order) stationary via hypothesis testing. We employed two tests for stationarity-- a Fourier-based test (the Priestley-Subba Rao (PSR) \citep{priestley1969test} test) and a wavelet-based test \citep{nason2013test}. The results (Table \ref{tab:stattest} in Appendix \ref{App:RDASupMat}) show that, for each of our motivating example datasets, over 80\% of the time series provided enough evidence to reject the null hypothesis of stationarity. This result suggests that the application of the current methodology (which assumes data stationarity) would be inappropriate for our motivating datasets and highlights the urgent need for more statistically advanced approaches.

The primary contribution of this work is the development of novel wavelet-based hypothesis tests that allow for circadian behaviour comparison while accounting for data nonstationarity. A substantial body of circadian literature advocates the use of wavelet \citep{price2008waveclock,harang2012wavos, leise2013wavelet} and in particular spectral representations \citep{hargreaves17:cluster} of circadian rhythms. This motivates our choice to formally compare circadian signals in the wavelet spectral domain by using their time-scale signature patterns and thus accounting for their proven nonstationary features.

This article is organised as follows. Section \ref{sec:lit review} reviews the theoretical wavelet-based framework we adopt for modelling nonstationary data and the relevant literature on hypothesis testing in the spectral domain. Our new hypothesis testing procedures are introduced in Section \ref{sec:hyptest}. Section~\ref{sec:sims} provides a comprehensive performance assessment of our new methods via simulation. Section~\ref{sec:real1} demonstrates the additional insight our techniques provide for the motivating circadian datasets and Section \ref{sec:concs} concludes this work.

\section{Overview: nonstationary processes and hypothesis testing in the spectral domain}
\label{sec:lit review}

\subsection{Modelling nonstationary processes}\label{sec:lsw}
Many of the statistically rigorous approaches to modelling nonstationary time series stem from the Cram\'er-Rao representation of stationary processes that states that all zero-mean discrete time second-order stationary time series $\{X_t\}_{t \in \ \mathbb{Z}}$ can be represented as
\begin{equation}
  \label{eq:SRT}
  X_t = \int_{-\pi}^{\pi} A(\omega)\exp(i\omega t ) d\xi(\omega),
  \end{equation}
where $A(\omega)$ is the amplitude of the process and $d\xi(\omega)$ is an orthonormal increments process \citep{priestley1983spectral}. In the representation of stationary processes (equation \eqref{eq:SRT}), the amplitude $A(\omega)$ does not depend on time, i.e.\ the frequency behaviour is the same across time. However, for many real time series, including our circadian datasets, this is unrealistic \citep{price2008waveclock} and a model where the frequency behaviour can vary with time is needed. Motivated by literature advocating wavelets as analysis tools for circadian rhythms \citep{leise2013wavelet}, we adopt the locally stationary wavelet (LSW) process model of \cite{nason2000wavelet} with previously demonstrated utility for circadian analysis \citep{hargreaves17:cluster}. In a nutshell, the Fourier building blocks in equation \eqref{eq:SRT} are replaced by families of discrete nondecimated wavelets and an LSW process $\{X_{t;T}\}_{t=0}^{T-1}$, $T=2^J \geq 1$ is represented as follows
\begin{equation}
  \label{LSW rep}
  X_{t,T} = \sum_{j = 1}^{J} \sum_{k \in \Z} w_{j, k;T} \psi_{j,k}(t)\xi_{j,k},
\end{equation}
where $\{\xi_{j,k}\}$ is a random orthonormal increment sequence, $\{\psi_{j, k}(t) = \psi_{j, k-t} \}_{j,k}$ is a set of discrete non-decimated wavelets and $\{ w_{j, k;T} \}$ is a set of amplitudes, each of which at a scale $j$ and time $k$. In this paper, we assume the $\xi_{j,k}$ are normally distributed (as in \cite{fryzlewicz2009consistent} and \cite{nason2015bayesian}). The properties of the random increment sequence $\{\xi_{j,k}\}$ ensure that $\{X_{t,T}\}$ is a zero-mean process. In practice it is customary to estimate and subtract the mean of a process with non-zero mean, and this is our approach here. A number of smoothness assumptions on the $\{ w_{j, k;T} \}_{j,k}$ are also required to allow estimation, see \cite{nason2000wavelet} for details.

Under the LSW framework, a quantity analogous to the spectrum of a stationary process is the evolutionary wavelet spectrum (EWS) $S_j(z) \coloneqq |W_j (z)|^2,$ at each scale $j \in \overline {1, J}$ and rescaled time $z=k/T \in (0,1)$. The EWS quantifies the power distribution in the process over time and scale. We define the raw wavelet periodogram as $I_{j,k;T} = \lvert d_{j,k;T} \rvert^2$, where $d_{j,k;T} = \sum_{t=0}^T X_{t, T} \psi_{j,k}(t)$ are the empirical nondecimated wavelet coefficients. In the remainder of this paper we drop the explicit dependence on $T$ for the wavelet coefficients and the periodogram.

The raw wavelet periodogram is an  asymptotically unbiased estimator of the quantity
\begin{equation}
\label{eq:beta}
\beta_j (z) = \sum_{i=1}^J A_{i, j} S_i (z)=(AS)_j(z),
\end{equation}
where $A=(A_{i,j})_{i,j=1}^J = (\sum_\tau  \Psi_i(\tau)\Psi_j(\tau))_{i,j=1}^J$ is the autocorrelation wavelet inner product matrix, with $\Psi_j (\tau) = \sum_k \psi_{j,k}(0) \psi_{j,k} (\tau)$ the autocorrelation wavelet \citep{nason2000wavelet}. The quantity $\beta_j (z)$ was introduced by \cite{fryzlewicz2006haar} and is often easier to work with theoretically than the spectrum \citep{nason2013test}. An asymptotically unbiased estimator of the EWS is the empirical wavelet spectrum:
\begin{equation}
\label{eq:corrected per}
 \mathbf{L}(z) \coloneqq A^{-1} \mathbf{I}(z),
\end{equation}
 $\mbox{ for all } z \in (0,1)$, where $\mathbf{I}(z):= (I_{j,[zT]})_{j=1}^J$  is the raw wavelet periodogram vector. 

The empirical wavelet spectrum is a collection of random variables that are not independent, nor is their (joint or marginal) distribution easy to determine. As each coefficient of the empirical wavelet spectrum is a sum of a (typically logarithmic) number of terms (see equation \eqref{eq:corrected per}), a central limit theorem-type mechanism brings it closer to normality than the raw wavelet periodogram \citep{fryzlewicz2009consistent}, which is distributed as a scaled $\chi^2_1$. As the individual raw periodogram ordinates within each scale are correlated,
\cite{fryzlewicz2006haar} model the raw wavelet periodogram as
\[
I_{j,k} \sim \beta_j(z) Z_{j, k}^2,
\]
where $z = k/T$ and $Z_{j, k}^2 \sim \chi_1^2$, for $j \in \mathbb{N}, \, k = 0, \dots, 2^J - 1 = T - 1$.

A way to `correct' these undesirable features is to employ a transform that brings the raw periodogram ordinates closer to Gaussianity and decorrelates within each scale. We adopt the Haar-Fisz transform (denoted $\mathcal{F}$), introduced (for spectral estimation) by \cite{fryzlewicz2006haar} and apply it separately to each scale $j = 1, \dots, J$ of the raw wavelet periodogram, denoted $\mathcal{H}_{j,k;T} \coloneqq \mathcal{F} I_{j,k;T}$. Proposition 6.1 in \cite{fryzlewicz2006haar} then suggests a potential model
$$
\mathcal{H}_{j,k} \sim N(\mathcal{B}_{j}(z), \sigma_j^2),
$$
where $\mathcal{B}_j(z) = \mathcal{F} \beta_j(z) \mbox{ with } z = k/T$ and $\mathcal{F} Z_{j, k}^2 \sim N(0,\sigma_j^2)$ and again dropping the explicit dependence on $T$. This model, viewed as a nonparametric additive regression model, was also employed by \cite{nason2015bayesian} in the context of Bayesian spectral estimation, where its viability was demonstrated.

\subsection{Spectral domain hypothesis testing}
\label{subsec:LitReviewHypTests}
Assuming that the available data consists of multiple nonstationary time series with known group memberships, to the authors' knowledge no hypothesis tests exist to determine whether two groups are significantly different in terms of their associated (evolutionary) wavelet spectra. The closest methodology for spectral comparison is framed as a (consistent) classification method of \cite{fryzlewicz2009consistent}, further improved by \cite{krzemieniewska2014classification}. \cite{tavakoli2016detecting} compare pairs of stationary functional time series, by developing $t$-tests for the equality of  their (Fourier) spectral density operators. \cite{shumway1988applied} compares groups of curves (with stationary stochastic errors) by testing whether the mean curves have the same Fourier spectrum at each given frequency. \cite{fan1998test} developed this method by applying the adaptive Neyman test to the (Fourier or wavelet) transformed difference  vector (the difference between the two group-average time series).  \cite{mckay2012statistically} developed wavelet-based functional ANOVA (wfANOVA) as an approach for comparing neurophysiological signals that are functions of time. This approach was also adopted by \cite{atkinson2017wavelet} to develop an approach to model validation using a test statistic based on thresholded wavelet coefficients. The emphasis of our work is different in that we combine the use of wavelets with rigorous stochastic nonstationary time series modelling.

Spectral comparison, framed as testing for spectral constancy, also appears in connection with testing for time series stationarity and white noise testing. In the Fourier domain, \cite{priestley1969test} determined (as a hypothesis test) whether the spectrum is time-varying and, hence, whether the process is nonstationary. \cite{von2000wavelet} introduced the principle of assessing the constancy of the time-varying Fourier spectrum by examining its Haar wavelet coefficients across time. In the wavelet domain, \cite{nason2013test} developed a test for second-order stationarity which examines the constancy of a wavelet spectrum by also examining its Haar wavelet coefficients. A similar approach is adopted by \cite{nason2014white} in the development of white noise tests.

\section{Proposed spectral domain hypothesis tests}\label{sec:hyptest}

Aligned to our motivating examples, the key goals of our work are to develop novel hypothesis tests, each capable of detecting one of three specific types of spectral differences between two groups and to identify the scales and times (e.g. Lead and Nematode datasets-- Sections \ref{subsubsec:Lead Intro}  and \ref{subsubsec:Worm Intro}) or scales only (e.g. Ultradian dataset-- Section \ref{subsubsec:AM Intro}) at which these difference arise, as appropriate.

Formally, we model the observed nonstationary circadian rhythms using the LSW framework of \cite{nason2000wavelet} (see Section \ref{sec:lsw} for details). Denote each individual profile by $\{ X_{t,T}^{(i),r_i}\}_{t=0}^{T-1}$ with $i=1, \, 2$ corresponding to one of two groups (e.g. control/ treatment) and potential replicates $r_i=1, \ldots, N_i$ (i.e. $N_i$ circadian traces in the $i$th group). Note that when $N_i=1$ we drop the $r_i$ index for simplicity.
Assume the signals in group $i$ are underpinned by a common wavelet spectrum and denote this by $S_{j}^{(i)}(t/T)$ for each group $i=1, \, 2$ at scales $j \in \overline {1, J}$ ($J = \log_2 T$) and rescaled times $z=t/T\in (0,1)$.

\subsection{Lead dataset: Hypothesis testing for spectral equality (`WST' and `FT')}\label{sec:wfst}

Put simply, our soil pollutant example focussed on detecting whether the two plant groups, `Control' and `Lead', display significant differences in the evolution of their spectral structures, and if so, the particular scales and times at which such differences occur. Mathematically we formalise our hypotheses as
\beq\label{eq:equal}
H_0: S_j^{(1)} (z) = S_j^{(2)} (z), \quad \forall j, \, z
\eeq
versus the alternative $H_A: S_{j^*}^{(1)} (z^*) \neq S_{j^*}^{(2)} (z^*)$ for some scale $j^*$ and rescaled time $z^*$. In the time domain, we visually note that differences in the circadian rhythms of the two groups appear towards the end of the experiment (see Figure \ref{fig:TSLeadpdf}).

\subsubsection{A naive wavelet spectrum test (`WST')}
\label{subsubsec:WST}
Since in reality we do not know the group spectrum $S_j^{(i)} (z)$, we replace it with a well-behaved estimator, denoted $\hat{S}_j^{(i)} (z)$. Assuming independent replicates are available for each group, we use the group ($i=1, \, 2$) averaged spectral estimators
\begin{equation}
\label{eq:sample coef est}
\hat{S}_j^{(i)} (k/T) = \frac{1}{N_i} \sum_{r_i=1}^{N_i}L_j^{(i), r_i} (k/T),
\end{equation}
where $L_j^{(i), r_i} (k/T)$ is the empirical wavelet spectrum of the $r_i$th series in group $i$ at scale $j$ and time $k$.  Should our spectral estimators satisfy the classical assumptions for a $t$-test (which in our context amount to independence of the spectral estimates across replicates and a Gaussian distribution), we propose a naive {\em wavelet spectrum test} (WST), centred on a test statistic of the form
\begin{equation}
\label{eq:test stat}
T_{j, k} = \frac{\hat{S}_j^{(1)} (k/T) - \hat{S}_j^{(2)} (k/T)}
{\left((\hat{\sigma}_{j, k}^{(1)})^2/N_1 + (\hat{\sigma}_{j, k}^{(2)})^2/N_2\right)^{1/2}} \sim t_{df} \mbox{ under the null hypothesis,}
\end{equation}
where $(\hat{\sigma}_{j, k}^{(i)})^2$ is an estimate of the variance of $\hat S_j^{(i)} (k/T)$ for $i = 1,2$ across the $N_i$ observations in group $i$, obtained using the standard sum--of--squares sample variance formula (as in \cite{krzemieniewska2014classification}). Under the null hypothesis of spectral equality, $T_{j,k}$ has a $t$-distribution with the number of degrees of freedom ($df$) directly related to the variance estimation procedure we employ. Each test statistic is then compared with a critical value derived from the $t$-distribution in the usual way.

When the variance of $\hat S_j^{(i)} (k/T)$ is unknown but common to both $i=1, \, 2$ groups (denoted $(\sigma_{j, k})^2 \coloneqq (\sigma_{j, k}^{(1)})^2 = (\sigma_{j, k}^{(2)})^2 $), it can be estimated using the pooled estimator:
\begin{equation}
\label{eq:pooled sd}
\hat{\sigma}_{j, k}^2 = \frac{(N_1 - 1)(\hat{\sigma}_{j, k}^{(1)})^2 + (N_2 - 1)(\hat{\sigma}_{j, k}^{(2)})^2}{N_1 + N_2 - 2},
\end{equation}
replacing $(\hat{\sigma}_{j, k}^{(1)})^2$ and $(\hat{\sigma}_{j, k}^{(2)})^2$ in equation~\eqref{eq:test stat}. The number of degrees of freedom in the $t$-distribution of the test statistic is then $df=N_1+N_2-2$.

If there is no reason to believe the group variances are equal, then use a $t$-distribution with degrees of freedom
\[
df=\frac{\left((\hat{\sigma}_{j, k}^{(1)})^2/N_1+
(\hat{\sigma}_{j, k}^{(2)})^2/N_2\right)^2}
{\frac{\left(
(\hat{\sigma}_{j, k}^{(1)})^2/N_1\right)^2}{N_1-1}+\frac{\left(
(\hat{\sigma}_{j, k}^{(2)})^2/N_2\right)^2}{N_2-1}}.
\]
However, the test statistic does not exactly follow the $t$-distribution, since two standard deviations are estimated in the statistic. Conservative critical values may also be obtained by using the $t$-distribution with $N$ degrees of freedom, where $N$ represents the smaller of $N_1$ and $N_2$ \citep{moore2007basic}.

In practice, the spectral estimators in equation \eqref{eq:sample coef est} may breach the Gaussianity testing assumption, especially when only a low number of replicates are available. The assumption of approximate normality for individual replicate spectral estimates, cautiously used in \citep{fryzlewicz2009consistent}, will be strengthened by the presence of a higher collection of group replicates ($N_1$, $N_2$) (see Section \ref{sec:sims} for a discussion of WST's features and caveats). In situations where the number of replicates is small, extra smoothing over time (e.g., via kernels) may be required \citep{fryzlewicz2009consistent}.

\subsubsection{Raw periodogram F-Test (`FT')}
\label{subsubsec:FT}
We now construct a testing procedure that is not reliant on the Gaussianity assumption whose validity we challenged above. Formally, for each scale $j \in \mathbb{N}$ and rescaled time $z \in (0, 1)$, the spectral equality $S_j^{(1)} (z) = S_j^{(2)} (z)$ is equivalent to  $\beta_j^{(1)} (z) = \beta_j^{(2)} (z)$ as the autocorrelation wavelet inner product matrix $A$ that links the two (see equation~\eqref{eq:beta}) is invertible. We therefore replace our initial collection of multiple hypothesis tests with equivalent re-framed versions
\[
H_0: \beta_j^{(1)} (z)= \beta_j^{(2)} (z), \forall j, \, z
\]
against the alternative that there exist a scale $j^*$ and rescaled time $z^*$ such that
$H_A:\beta_{j^*}^{(1)} (z^*) \neq \beta_{j^*}^{(2)} (z^*)$. In order to construct our test statistic, we test for spectral equality by examining the $\beta_j (z)$ quantities instead.

In reality we do not know $\beta_j^{(i)} (z)$ for $i=1,\, 2$ so we replace it by an asymptotically unbiased estimator. As data are available consisting of multiple time series with known group memberships, we replace $\beta_j^{(i)} (z)$ with an estimate across the group replicates. Specifically, if we have $N_i$ independent time series replicates from group $i$, we define
\begin{equation}
\label{eq: FT test stat sum}
N_i \bar{I}_{j, k}^{(i)} \coloneqq \sum_{r_i = 1}^{N_i} I_{j, k}^{(i), r_i} \sim \chi^2_{N_i}(\beta_j^{(i)} (k/T)).
\end{equation}

The distribution above follows as the raw wavelet periodogram coefficient of each $r_i$th periodogram replicate $I_{j, k}^{(i), r_i}$ is (scaled) $\chi_1^2$ distributed (e.g. \cite{nason2015bayesian}) and independent of all other raw wavelet periodogram coefficients across all other replicates from the same group (also see \cite{fryzlewicz2009consistent} and the discussion in Section \ref{sec:lsw}). Under the further assumption of group independence, $\bar{I}_{j, k}^{(1)}$ and $\bar{I}_{j, k}^{(2)}$ are independent and distributed as detailed in equation~\eqref{eq: FT test stat sum}. Hence we propose the test statistic
\begin{equation}
\label{eq:FT test stat}
F_{j, k} = \frac{\bar{I}_{j, k}^{(1)} } {\bar{I}_{j, k}^{(2)} } \sim F_{N_1, N_2} \mbox{ under the null hypothesis.}
\end{equation}
Each test statistic is then compared with a critical value derived from the $F_{N_1, N_2}$-distribution in the usual way.

\noindent {\em Discussion.} An advantage of the FT, particularly as opposed to the WST,
is that its underlying distributional assumption is theoretically, as well as practically,
more reliable. We would therefore expect the FT to outperform the WST
in many applications, and this is indeed validated across a variety of simulation
settings (see Section \ref{sec:sims}).

As we wish to test many hypotheses of the type $H_0: \beta_j^{(1)} (k/T)= \beta_j^{(2)} (k/T)$ for several values of $j$ and $k$, we are in the field of multiple-hypothesis testing. For all tests we develop, we use Bonferroni correction and, for a less conservative approach, the false discovery rate (FDR) procedure introduced by \cite{benjamini1995controlling}. Our simulations in Section \ref{sec:sims} show that both these methods work well. However, of course the tests themselves are related to one another, but just as in \cite{nason2013test} we do not pursue this topic further in this work.

The WST and FT developed above both report the time-scale locations of the significant differences between the two group spectra. These can be visualised as a `barcode' plot, where a significant difference is represented by a black line at the time-scale location of the rejection of the null hypothesis (see for example Figure \ref{fig:AveSpecandHFTLead}, right). Alternatively, for all our proposed tests, practitioners can also be informed by the number of rejections (as a dissimilarity measure), with larger values indicating a greater departure from the null hypothesis  (as discussed in \cite{das2016measuring} and in Section \ref{subsec:sim size comparison}).

\subsection{Ultradian dataset: Hypothesis testing for spectral equality across scales (`HFT')}\label{sec:hft}

For certain biological applications, such as the Ultradian motivating example, it is more important to identify spectral differences between groups at scale-level and the time locations of spectral differences are of less interest. For such situations, we replace the spectral comparison $H_0: S_j^{(1)} (z) = S_j^{(2)} (z)$ of the previous section, in general equivalent to $H_0: \beta_j^{(1)} (z)= \beta_j^{(2)} (z)$, by the comparison of the respective Haar-Fisz transforms, i.e. test for
\[
H_0: \mathcal{F}\beta_j^{(1)} (z) = \mathcal{F}\beta_j^{(2)} (z), \forall j, \, z.
\]
Equivalently, in the notation established in Section \ref{sec:lsw} we test
\begin{equation}
\label{eq:HFT hyps}
H_0: \mathcal{B}_{j}^{(1)}(z) = \mathcal{B}_{j}^{(2)} (z), \, \forall j, \, z
\end{equation}
versus the alternative that there exist some scale $j^*$ and rescaled time $z^*$ for which the equality does not hold. We shall refer to this test as the {\em Haar-Fisz test} (HFT). Intuitively, although the HFT identifies both scales and times at which the null hypothesis of spectral equality in the Haar-Fisz domain does not hold, as the Haar-Fisz transform essentially `averages' within each scale of the raw wavelet periodogram, potential differences  `spread' throughout the scale. This property makes it ideal for identifying scale-level differences between group wavelet spectra (see for example Figure \ref{fig:AveSpecandHFT}, right).

As we do not know $\mathcal{B}_j^{(i)} (z)$, we replace it by its  unbiased estimator $\mathcal{H}_{j,k}^{(i)}$ at scale $j$ and time $k$ (with $z=k/T$) for group $i=1, 2$.  In applications which do not provide access to replicate data, we could adopt equation \eqref{eq:test stat} with $\hat{S}_j^{(i)} (k/T)$ replaced by $\mathcal{H}_{j,k}^{(i)}$ and estimate the variance across each scale as the Haar-Fisz transform stabilises variance \citep{nason2015bayesian} (see Appendix \ref{App:SimStudyDetails}). When replicates are available, we use equation \eqref{eq:sample coef est} with $\mathcal{H}_{j,k}^{(i)}$ to obtain group averaged estimators of $\mathcal{B}_j^{(i)} (z)$, denoted $\hat{\mathcal{H}}_{j,k}^{(i)}$, and propose a test statistic as in equation~\eqref{eq:test stat} with $\hat{S}_j^{(i)} (k/T)$ replaced by $\hat{\mathcal{H}}_{j,k}^{(i)}$. The variance estimation techniques and subsequent test statistic distribution follow as detailed in Section \ref{sec:wfst} and the results of the HFT can also be visualised as a `barcode' plot.

The rationale of this approach is to also bring the data (in this context, the Haar-Fisz transform of the raw wavelet periodogram) closer to Gaussianity and to break the dependencies across time. Consequently, the assumptions behind the $t$-test are closely adhered to and the dependencies between the multiple tests we perform are weak. In practice, due to its scale averaging construction, the HFT unsurprisingly results in many more time-localised rejections than the actual number of differing coefficients in the original spectra, and does sometimes have difficulty discriminating between spectra which differ by a small number of coefficients; however, the HFT does correctly identify scale-level spectral differences (see Section \ref{sec:sims} for further investigations).

\subsection{Nematode dataset: Hypothesis testing for `same shape' spectra (`HT')}\label{sec:ht}
In applications such as the Nematode example, the focus may be on identifying whether groups evolve according to spectra that have the same shape at each scale, thus indicating that the same patterns are identified in the data, albeit with potentially different magnitudes.

Mathematically, for a scale-dependent (non-zero) constant denoted by $C_j$, we formalise our hypotheses as
\beq\label{eq:const}
H_0: S_j^{(1)} (z) = S_j^{(2)} (z) + C_j, \quad \forall j, \, z
\eeq
versus the alternative $H_A: S_{j^*}^{(1)} (z^*) \neq S_{j^*}^{(2)} (z^*) + C_{j^*}$ for some scale $j^*$ and time $z^*$.

Denoting by $\underline{C}$ the $J \times 1$ vector that holds $C_j$ as its $j$th component and recalling equation \eqref{eq:beta}, we can equivalently re-frame the problem into testing whether
$$H_0: \beta_j^{(1)} (z) = \beta_j^{(2)} (z) + c_j, \mbox{ or equivalently } H_0: \beta_j^{(D)} (z)=c_j, \quad \forall j, z$$ where $c_j$ is the $j$th entry of the vector $\underline{c}=A \underline{C}$ and $\beta_j^{(D)} (z) \coloneqq \beta_j^{(1)} (z) - \beta_j^{(2)} (z)$.

In the spirit of the tests developed in \cite{fan1998test}, and as undertaken by \cite{von2000wavelet} and  \cite{nason2013test}, at each scale $j$ we assess the constancy through time of $\beta_j^{(D)} (z)$ by examining its associated Haar wavelet coefficients. Although, in principle, any wavelet system could be adopted, \cite{von2000wavelet} note that the Haar wavelet coefficients are ideal for testing the constancy of a function. Hence we employ these wavelets and refer to the test developed in this section as the \textit{Haar Test} (HT).

The underlying principle behind these tests is that the wavelet transform of a constant function is zero, hence under $H_0$ above, the wavelet coefficients of $\beta_j^{(D)} (z)$ are
 \[
 v_{\ell, p}^{j} = \int_0^1 \beta_j^{(D)} (z) \psi_{\ell, p}^H (z)dz = c_j \int_0^1 \psi_{\ell, p}^H (z)dz = 0,
 \]
where $\{\psi_{\ell, p}^H (z)\}_{\ell, p}$ denote the usual Haar wavelets at scale $\ell$ and location $p$.

This suggests performing multiple hypothesis testing on the collection of hypotheses
\[
H_0: v_{\ell, p}^{j} = 0, \, \forall j, \ell \mbox{ and }p
\]
against the alternative that there exist $j^*, \ell^*$ and $p^*$ such that $H_A: v_{\ell^*, p^*}^{j^*} \neq 0$. 

As the spectral and related quantities are unknown, and since the wavelet transform is linear, we estimate each $v_{\ell, p}^{j}$ by $\hat{v}_{\ell, p}^{j} = \hat{v}_{\ell, p}^{j, (1)} - \hat{v}_{\ell, p}^{j, (2)}$, with the Haar wavelet coefficients corresponding to each  group $i=1, \,2$ estimated in the spirit of \cite{nason2013test} as
\begin{equation}
\label{eq: Haar Wav coeffs}
\hat{v}_{\ell, p}^{j, (i)} = 2^{-\ell/2} \Bigg(\sum_{r=0}^{2^{\ell-1}-1} I_{j, 2^\ell p-r}^{(i)} - \sum_{q=2^{\ell-1}}^{2^{\ell}-1} I_{j, 2^\ell p-q}^{(i)}\Bigg),
\end{equation}
at each (original) scale $j$ and Haar scale $\ell$ and locations $p$, $q$. 

With the availability of independent replicates within each group, we estimate the group $i$ Haar wavelet coefficients as
\begin{equation}
\label{eq:HT average}
\hat{v}_{\ell, p}^{j, (i)} = \frac{1}{N_i} \sum_{r_i=1}^{N_i}\hat{v}_{\ell, p}^{j, (i), r_i},
\end{equation}
where each $\hat{v}_{\ell, p}^{j, (i), r_i}$ is obtained as in equation \eqref{eq: Haar Wav coeffs} for the $r_i$-th replicate.

Under a specific set of assumptions, \cite{nason2013test} shows the asymptotic normality of the Haar wavelet coefficient estimator of the wavelet periodogram at scale $j$. Thus, in our setting, each $\hat{v}_{\ell, p}^{j, (i), r_i}$ for $i=1, \, 2$ is asymptotically normal with mean ${v}_{\ell, p}^{j, (i), r_i}$ and variance $({\sigma}_{\ell, p}^{j, (i)})^2$. Using the replicate independence, we have that $\hat{v}_{\ell, p}^{j, (i)}$ is asymptotically normally distributed with mean $v_{\ell, p}^{j, (i)}$ and variance $({\sigma}_{\ell, p}^{j, (i)})^2/N_i$ and note that its distributional closeness to the normal increases via a central limit theorem argument with the increasing number of replicates.

The group independence assumption then leads to an asymptotically joint normal distribution for $(\hat{v}_{\ell, p}^{j, (1)},\hat{v}_{\ell, p}^{j, (2)})$. Following the continuous mapping theorem, we obtain that 
$\hat{v}_{\ell, p}^{j} = \hat{v}_{\ell, p}^{j, (1)} - \hat{v}_{\ell, p}^{j, (2)}$ has an asymptotic normal distribution with mean ${v}_{\ell, p}^{j, (1)} - {v}_{\ell, p}^{j, (2)}$ and variance $\left(({\sigma}_{\ell, p}^{j, (1)})^2/N_1 + ({\sigma}_{\ell, p}^{j, (2)})^2/N_2 \right)$.

In the presence of replicates, we propose a test statistic of the form discussed in equation~\eqref{eq:test stat}

\begin{equation}
\label{eq:HT test stat pooled}
T_{\ell, p}^{j} = \frac{\hat{v}_{\ell, p}^{j}}
{\left((\hat{\sigma}_{\ell, p}^{j, (1)})^2/N_1 + (\hat{\sigma}_{\ell, p}^{j, (2)})^2/N_2 \right)^{1/2}} \sim t_{df} \mbox{ under the null hypothesis,}
\end{equation}
where $(\hat{\sigma}_{\ell, p}^{j, (i)})^2$ is an estimate of the variance of $\hat{v}_{\ell, p}^{j, (i)}$ for $i = 1,2$ across the $N_i$ observations in group $i$, obtained using the standard sum--of--squares sample variance formula and $df$ denotes the degrees of freedom associated with the variance estimation procedure (see Section \ref{subsubsec:WST}). Each test statistic is then compared with a critical value derived from the $t$-distribution in the usual way. 

In order to control the asymptotic bias derivation, one of the assumptions under which the distributional theory is derived consists of limiting the scales of the Haar wavelet coefficients $v_{\ell, p}^{j}$ to be sufficiently coarse, $\ell= 0, \dots, (J - \lceil J/2 \rceil -2)$. Furthermore, as in \cite{nason2013test}, we only consider the wavelet coefficients of the periodogram at levels $j \geq 3$ in order to avoid the effects of a region similar to the `cone of influence' described by \cite{torrence1998practical}.


To aid the visualisation of the WST, FT and HFT results, we use a `barcode' plot that indicates the time- and scale- locations where significant differences are present. The HT can also indicate where the significant differences are located in the series and can plot the results in a manner similar to the wavelet test of stationarity (see \cite{nason2013test}). However, due to its construction, these locations are more difficult to interpret than for the WST, FT and HFT (see Figure \ref{fig:WormsSpecandTests2}).

\section{Simulation Studies}\label{sec:sims}
The goals of the simulation studies were: (1) to evaluate the empirical power and size of our new tests; (2) to consider the effect of sample size on the accuracy of the tests; (3) to investigate two approaches to multiple-hypothesis testing: Bonferroni correction (denoted `Bon.') and the false discovery rate procedure (denoted `FDR') and (4) to evaluate the empirical power and size of our new tests in comparison with the adaptive Neyman Test (ANT) of \cite{fan1998test}, see Section \ref{subsec:LitReviewHypTests}. This benchmark method performs well in practice when the assumption that the data can be modelled as a functional time series is valid. The ANT can also be conceptualised as an advanced version of the analysis a circadian biologist would currently perform.

In this section we briefly outline the basic structure of each simulated experiment (a comprehensive description of the simulation studies can be found in Appendix \ref{App:SimStudyDetails}). In each case, we assumed that the signal was a realisation from one of $i=1,2$ possible groups. For each group, we generated a set of $N_1 = N_2 = 1, 10, 25, 50$ signal realisations, each of length $T=256$, the equivalent of a free-running period of 4 days. For each realisation, we obtained the raw and corrected wavelet periodograms using the Haar wavelet from the \verb|locits| software package for R (available from the CRAN package repository), although, any wavelet system can, in principle be used (see Section \ref{sec:concs} for a discussion). The Haar--transformed and Haar-Fisz transformed raw wavelet periodogram were subsequently obtained and the spectral testing procedures carried out as described in Section \ref{sec:hyptest}. The results are compared with the known group memberships, and the procedure is then repeated $1000$ times to obtain empirical size and power estimates as outlined in the following sections.

\subsection{Power Comparisons}
\label{subsec:sim power comparison}
To explore statistical power we simulate a set of $N_1 = N_2 = 1, 10, 25, 50$ signal realisations from each group where the individual group spectra are defined such that there exists a scale $j^*$ and time $t^*$ such that $H_A: S_{j^*}^{(1)} (t^*/T) \neq S_{j^*}^{(2)} (t^*/T)$. The empirical power estimates are obtained by counting the number of times our tests reject the null hypothesis of spectral equality. The models we will use are denoted {\bf P1}--{\bf P12} respectively and are briefly described below. 
 \begin{enumerate}
 \item \textbf{P1: Fixed Spectra.}
 We follow \cite{krzemieniewska2014classification} and design the spectra of the two groups to differ at the finest level (resolution level 7) by $100$ coefficients.

 \item\textbf{P2: Fixed Spectra-Fine Difference.}
 We modify the model {\bf P1} such that the spectra of the two groups differ by only $6$ coefficients.

 \item \textbf{P3: Fixed Spectra-Plus Constant.}
 Modify the model {\bf P1} such that the spectra of the two groups differ by a constant in the finest resolution level.

\item \textbf{P4/P5: Gradual Period Change.}
This study replicates a typical circadian experiment with changes that cannot be captured by standard analyses assuming stationarity and only reporting an average period value. We thus define $3$ possible groups, where each group represents a signal that gradually changes period from 24 to: 25 (Group 1), 26 (Group 2) and 27 (Group 3) over (approximately) two days, before continuing with the relevant period for a further two days (also see \cite{hargreaves17:cluster}). To determine which changes can be discriminated by the methods, we perform two studies within this setting: simulations from Groups 1 and 2 (\textbf{P4}) and simulations from Groups 1 and 3 (\textbf{P5}).

\item \textbf{P6/P7: AR Processes with time-varying coefficients.}
We simulate from an important class of nonstationary processes-- AR(2) processes with: abruptly (\textbf{P6}) and slowly (\textbf{P7}) changing parameters (as in \cite{fryzlewicz2009consistent}).

\item \textbf{P8--P12: Functional Time Series (Constant Period).} This study follows \cite{zielinski2014strengths} and generates each time series using an underlying cosine curve with additive noise, which also coincides with the theoretical assumptions of the ANT. We define time series as realisations from one of 6 possible groups, each with a different (constant) period, relevant to our circadian setting. To determine which period changes can be discriminated by the methods, we perform five studies within this setting: simulations from a group with a period of 24 hours versus a group with a period of 21, 22, 23, 23.5 and 23.75 hours (models \textbf{P8--P12} respectively). 

\end{enumerate}

\noindent{\bf Discussion of findings.}
The empirical power values for $N_1 = N_2 = 25$ (this is the typical number of available replicates in circadian studies, see Section \ref{sec:real1}) for models \textbf{P1--P7} are reported in Table \ref{tab:sim25}. We found that all tests perform well when the spectra differ by a large number of coefficients (model \textbf{P1}). The FT (and, to a lesser extent, the HT) are able to discriminate between spectra that differ by a small number of coefficients (model \textbf{P2}) whereas the HFT has lower empirical power. By construction, the HT cannot differentiate between spectra that differ by a constant at a particular resolution level (model \textbf{P3}), but we found that the HT performs well in our synthetic circadian example of gradual small period change across many time-scale locations (models \textbf{P4} and \textbf{P5}). Due to the higher distributional reliability of the FT, it unsurprisingly outperforms the WST when the times series are generated from a defined spectrum (models \textbf{P1}--\textbf{P5}). However, distributional properties of the time-varying AR process ensure that the WST performs best when data are generated using models \textbf{P6} and \textbf{P7}, with the HT and HFT also performing well for model \textbf{P7}. We also report the empirical power of the ANT for model \textbf{P5} (gradual period change, 25 replicates) was $10.7\%$ which is below the results in Table \ref{tab:sim25} for our proposed tests. This is to be expected as the underlying assumptions of the ANT are no longer met (similar results are obtained for models \textbf{P1}--\textbf{P7}, hence we do not provide these here).

The numbers of replicates in each group ($N_1, N_2$) are also an important factor in power. The results for $N_1 = N_2 =1, 10$ and $50$ replicates are shown in Tables \ref{tab:simPower1}, \ref{tab:sim10} and \ref{tab:sim50} (Appendix \ref{App:SupTabs}) respectively. Increasing the number of replicates should, and indeed does, increase the empirical power of all tests (with the exception of the HT for model \textbf{P3}). For example, note the increase in empirical power (particularly, for models \textbf{P2} and \textbf{P4}) as the number of replicates increases from $10$ to $25$.

Table \ref{tab:sim_Comp} presents a selection of the performance comparison results for models {\bf P8--P12}. As expected, the ANT performs extremely well in all these studies since the underlying assumptions of the methodology are adhered to. Nevertheless, it is very pleasing that the WST, FT and HT also all have an empirical power over $95\%$ (25 replicates) showing that our methodology can also be successfully applied to functional time series as designed for the ANT. However, the HFT had difficulty discriminating between groups when the period difference was less than 2 hours. This was no surprise as the HFT was constructed to detect differences in scale only and, due to the lower frequency resolution of the wavelet spectrum, the total power within each scale of the wavelet spectrum will be very similar for both groups.


\begin{table}
\begin{center}
\begin{tabu}{|c|[2pt]c|c|[2pt]c|c|[2pt]c|c|[2pt]c|c|}
\hline \thead{Model} & \thead{WST \\ (Bon.)} & \thead{WST \\ (FDR)} & \thead{FT \\ (Bon.)} & \thead{FT \\ (FDR)} & \thead{HFT \\ (Bon.)} & \thead{HFT \\ (FDR)} & \thead{HT \\ (Bon.)} & \thead{HT \\ (FDR)} \\
\hline \textbf{P1} & 100.0 & 100.0 & 100.0 & 100.0 & 100.0 & 100.0 & 100.0 & 100.0 \\
\hline \textbf{P2} & 39.3 & 48.0 & \textbf{100.0} & \textbf{100.0} & 29.1 & 31.8 & 86.2 & 86.4 \\
\hline \textbf{P3} & 100.0 & 100.0 & 100.0 & 100.0 & 100.0 & 100.0 & 4.3 & 4.4 \\
\hline \textbf{P4} & 1.0 & 2.7 & 45.5 & 54.5 & 33.2 & 36.5 & \textbf{100.0} & \textbf{100.0} \\
\hline \textbf{P5} & 5.9 & 14.6 & 97.0 & 99.9 & \textbf{100.0} & \textbf{100.0} & \textbf{100.0} & \textbf{100.0} \\
\hline \textbf{P6} & \textbf{100.0} & \textbf{100.0} & 87.5 & 92.6 & 44.8 & 89.1 & 66.5 & 67.7 \\
\hline \textbf{P7} & \textbf{100.0} & \textbf{100.0} & 54.3 & 64.5 & 97.4 & 99.9 & \textbf{100.0} & \textbf{100.0} \\
\hline
\end{tabu}
\caption{Simulated power estimates ($\%$) for models P1-P7 with nominal size of $5\%$ with $N_1 = N_2 = 25$ realisations from each group. Highest empirical power estimates are highlighted in bold.}
  \label{tab:sim25}
\end{center}
\end{table}

 \subsection{Size Comparisons}
 \label{subsec:sim size comparison}
To explore statistical size, we simulate data from a number of models and we asses how often our hypothesis tests reject the null hypothesis of spectral equality (i.e. the time series are generated in the same way for both test groups). The models are denoted {\bf M1}--{\bf M5} respectively and defined as follows.
\begin{enumerate}
\item \textbf{M1: Fixed Spectra.}
We simulate all data from the wavelet spectrum associated with Group $1$ in models \textbf{P1}, \textbf{P2} and \textbf{P3}, which we define as  $\{ S_{j}^{(1)}(z) \}_{j=1}^{J}$ in equation \eqref{eq:KrezSim1}.
\item \textbf{M2: Gradual Period Change.}
We simulate all data from the wavelet spectrum which corresponds to a time series that gradually changes period from 24 to 25 hours over (approximately two days), before continuing with period 25 hours for a further two days (i.e. Group 1 from models \textbf{P4/P5}).
\item \textbf{M3: AR Processes With Abruptly Changing Parameters.}
Each time series is generated from the process defined by equation \eqref{eq:Case5.1}
with the abruptly changing parameters as defined for group $i=1$ in Table \ref{tab:5.1Bparams} (i.e. Group 1 from model \textbf{P6}).
\item \textbf{M4: AR Processes With Slowly Changing Parameters.}
Each time series is generated from the process defined by equation \eqref{eq:Case5.2} with the slowly changing parameters as defined for group $i=1$ in Table \ref{tab:5.2Bparams} (i.e. Group 1 from model \textbf{P7}).
\item \textbf{M5: Functional Time Series (Constant Period). } 
All data are simulated (using equation \eqref{eq:FuncTS}) from the model that corresponds to a time series with a constant period of 24 hours (i.e. Group 1 from models \textbf{P8--P12}).
\end{enumerate}

\noindent{\bf Discussion of findings.}
The empirical size values for models \textbf{M1--M4} with $N_1 = N_2 = 25$ (this is the typical number of available replicates in circadian experiments, see Section \ref{sec:real1}) are reported in Table \ref{tab:Sizesim25}. The results for $N_1 = N_2 =1, 10$ and $50$ are shown in Tables \ref{tab:simSize1}, \ref{tab:Sizesim10} and \ref{tab:Sizesim50} (Appendix \ref{App:SupTabs}) respectively. These studies show that the empirical size corresponding to all proposed tests (apart from the FT for model \textbf{M4} with $N_1 = N_2 = 10$ and $25$) are less than the nominal size of $5\%$, with the Bonferroni correction providing a more conservative approach and the false discovery rate being closer to the nominal size. A close inspection of rejections for the FT for model \textbf{M4} with $N_1 = N_2 = 10$ and $25$ and both multiple-hypothesis testing methods (Table \ref{tab:SimsCase 5.2B FT rejs} in Appendix \ref{App:SupTabs}) reveals that for this particular example the number of rejections is often 1. Thus, if we disregard such situations, the empirical size of the FT also falls below the nominal size of $5\%$ for all sample sizes and multiple-hypothesis testing procedures. In practice, circadian scientists are mostly interested in the numbers of rejections and their locations and often choose to disregard situations where very few coefficients are significantly different.  Indeed, this is also our approach in Section \ref{sec:real1}.  The results for model \textbf{M5} with $N_1 = N_2 = 10$ and $25$ are shown in Table \ref{tab:sim_Comp} (Appendix \ref{App:SupTabs}). Note that the empirical size estimates for our proposed tests are all lower than the nominal size of $5\%$, whereas for 10 replicates the empirical size of the ANT is $7.9\%$.


\begin{table}
\begin{center}
\begin{tabu}{|c|[2pt]c|c|[2pt]c|c|[2pt]c|c|[2pt]c|c|}
\hline \thead{Model} & \thead{WST \\ (Bon.)} & \thead{WST \\ (FDR)} & \thead{FT \\ (Bon.)} & \thead{FT \\ (FDR)} & \thead{HFT \\ (Bon.)} & \thead{HFT \\ (FDR)} & \thead{HT \\ (Bon.)} & \thead{HT \\ (FDR)} \\
\hline \textbf{M1} & 0.6 & 1.3 & 2.5 & 3.1 & 0.1 & 2.0 & 2.3 &  2.7\\
\hline \textbf{M2} & 0.3 & 0.6 & 3.0 & 3.9 & 0.4 & 3.3 & 2.5 & 2.7 \\
\hline \textbf{M3} & 0.2 & 1.5 & 3.6 & 3.9 & 0.0 & 1.6 & 3.5 & 3.8 \\
\hline \textbf{M4} & 0.4 & 0.9 & 4.6 & \textbf{5.2} & 1.0 & 2.4 & 3.4 & 3.8 \\
\hline
\end{tabu}
\caption{Simulated size estimates ($\%$) for models M1-M4 with nominal size of $5\%$ and $N_1 = N_2 = 25$ realisations from each group. Empirical size estimates over the nominal size of $5\%$ are highlighted in bold.}
  \label{tab:Sizesim25}
\end{center}
\end{table}


\section{Real data analysis: back to the motivating circadian datasets}
\label{sec:real1}
We now use our proposed methodology to analyse the motivating examples (Section \ref{sec:intro}).

As is typical for wavelet representations, the data is often required to be of dyadic length, $T = 2^J$ (see equation \eqref{LSW rep}). In many practical applications, this is not realistic and there are a number of approaches to address this situation (see e.g. \cite{todd1997preconditioning}). Our approach is to analyse a (dyadic length) segment of the data, with the truncation decided upon consultation with the experimental scientists. We then model each circadian trace as an LSW process, estimate its corresponding group wavelet spectral representation and consequently construct the appropriate test statistic that aims to identify whether a departure towards a specific type of spectral difference is present or not (as described in Section \ref{sec:hyptest}). For each dataset, the corresponding number of rejections can be found in Table \ref{tab:ResultsRDA} (Appendix \ref{App:RDASupMat}), with corresponding representative `barcode' plots in Figures \ref{fig:AveSpecandHFTLead}, \ref{fig:AveSpecandHFT} and \ref{fig:WormsSpecandTests2}.

\subsection{Lead dataset}
\label{subsec:app case 1 metals}
\begin{figure}
       \centering
       \includegraphics[width=\linewidth]{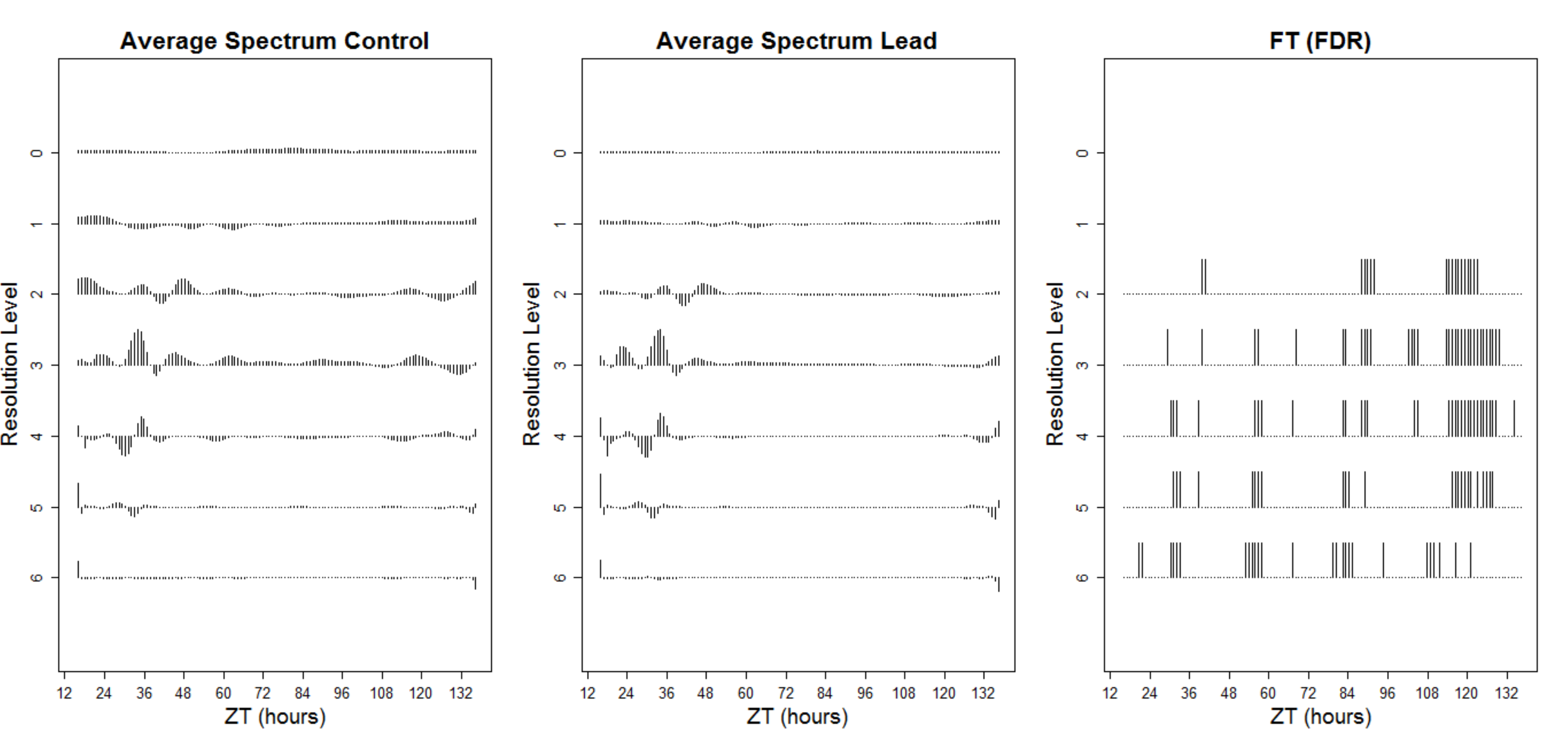}
       \caption{\textbf{Lead dataset.} Left: Average estimated spectrum of the `Control' group; Centre: Average estimated spectrum of the `Lead' group; Right: `Barcode' plot for FT (with FDR).}
       \label{fig:AveSpecandHFTLead}
       \end{figure}

Section \ref{subsubsec:Lead Intro} outlined the scientific aims to determine if lead nitrate affects the circadian clock and, if so, to detect the times and scales at which any significant differences arise between the `Control' and `Lead' exposure groups. Therefore we are particularly interested in the results of the FT. Table \ref{tab:ResultsRDA} shows the results for the FT and includes both the more conservative Bonferroni correction and FDR. In order to visualise the areas of null hypothesis rejection of spectral equality between the control and lead-exposure groups, both group average estimated spectra as well as the `barcode' plot for the FT (with FDR) appear in Figure \ref{fig:AveSpecandHFTLead}. Figure \ref{fig:AveSpecandHFTLead} indicates that the differences between the two spectra lie in resolution levels 2--4, directly corresponding to a circadian rhythm, with the number of rejections increasing with exposure time. We conclude that there is evidence that exposure to lead does affect the circadian clock of \textit{A. thaliana}, and this change manifests itself after approximately three days of free-running conditions.

\subsection{Ultradian dataset}
\label{subsec:app case 2 AM}
\begin{figure}
       \centering
       \includegraphics[width=\linewidth]{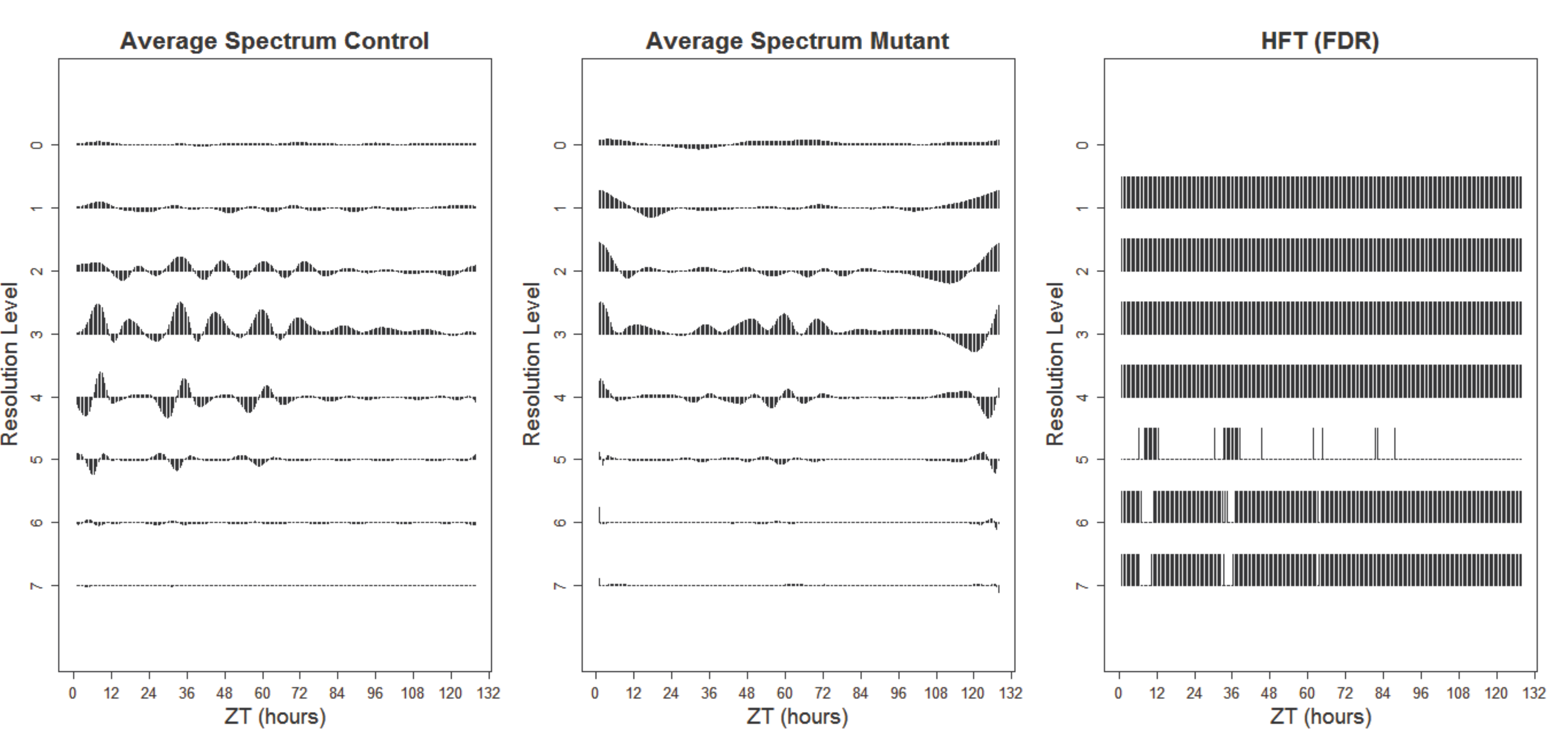}
       \caption{\textbf{Ultradian dataset.} Left: Average estimated spectrum of the `Control' group; Centre: Average estimated spectrum of the `Mutant' group; Right: `Barcode' plot for HFT (with FDR).}
       \label{fig:AveSpecandHFT}
       \end{figure}

Section \ref{subsubsec:AM Intro} introduced this experiment and highlighted the need to detect whether any differences appear
in the circadian and ultradian components of the `Control' and `Mutant' groups. Hence we are interested in the results of the HFT, specifically developed to identify the scales, rather than the times, at which potential differences arise. Table \ref{tab:ResultsRDA} shows the results for the HFT, including both the Bonferroni correction and FDR. The results indicate rejections of the null hypothesis of spectral equality between the control and mutant plants across a range of scales. The group average estimated spectra and `barcode' plot for the HFT (with FDR) can be found in Figure \ref{fig:AveSpecandHFT}. Note that the differences between the two spectra lie in the coarsest resolution levels 1--4, associated with circadian rhythms, and higher-frequency levels 6 and 7, corresponding to an ultradian rhythm. We conclude that there is evidence that the mutant plants have altered circadian and ultradian rhythms within \textit{A. thaliana}.

\subsection{Nematode dataset}
\label{subsec:app case 3 worms}
\begin{figure}
       \centering
       \includegraphics[width=\linewidth]{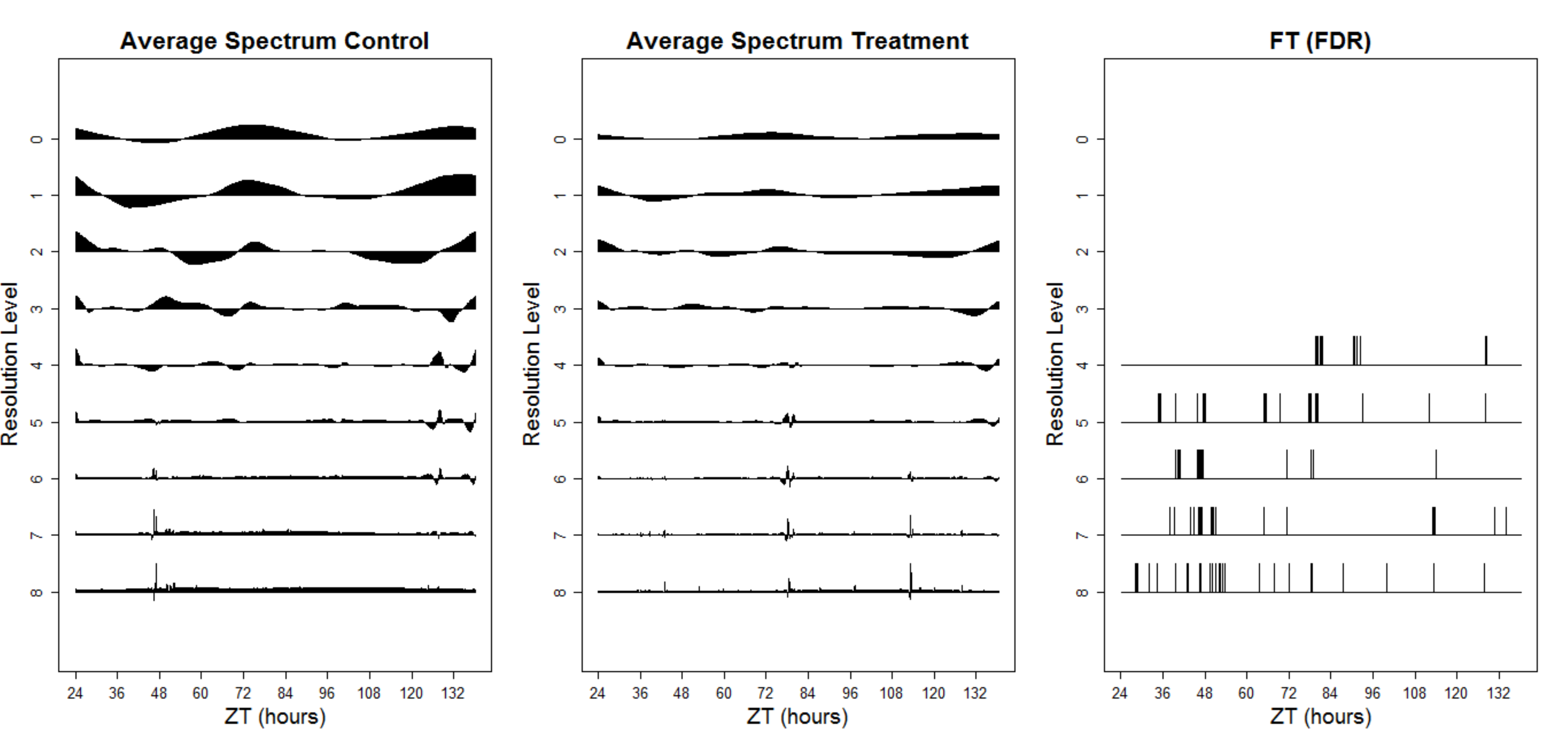}
       \caption{\textbf{Nematode dataset.} Left: Average estimated spectrum of the `Control' group; Centre: Average estimated spectrum of the `Treatment' group; Right: `Barcode' plot for FT (with FDR).}
       \label{fig:WormsSpecandTests2}
       \end{figure}

The experiment in Section \ref{subsubsec:Worm Intro} aimed to elucidate the effect of a pharmacological treatment on the {\em C. elegan} clock. The average estimated spectra of the `Control' and `Treatment' groups in Figure \ref{fig:WormsSpecandTests2} share a common profile but with differences in magnitude, indicating that the HT would be appropriate in this context. Table \ref{tab:ResultsRDA} shows that the HT found no significant difference between the shapes of the two spectra, but when tested for equality, the FT (with FDR) found multiple rejections of the null hypothesis of spectral equality between the `Control' and `Treatment' groups (refer to the `barcode' plot in Figure \ref{fig:WormsSpecandTests2}). This provides evidence that the two spectra have the same profile within each scale up to an additive non-zero constant. We thus conclude that there is evidence that the treatment significantly affects the intensity of the spectral behaviour, but not its pattern. The spectral differences are present at the highest frequencies (resolution levels 6--8) as an early response to the onset of treatment (prior to time $T=48$), see Figure \ref{fig:WormsSpecandTests2}.

\section{Conclusions and further work}
\label{sec:concs}

This work was stimulated by a variety of challenging applications faced by the circadian--biology community, which is becoming increasingly aware of the nonstationary characteristics present in much of their data \citep{hargreaves17:cluster, zielinski2014strengths, leise2013wavelet}. Our methodology fills the gap in the current literature by developing and testing a much needed tool for the formal spectral comparison of nonstationary data. Our methods are developed as testing procedures, analogous to the period analysis techniques currently adopted within circadian community. Their competitive performance was comparatively assessed in an extensive simulation study (Section \ref{sec:sims}). Additionally, when compared to existing methods that assume stationarity, our proposed tests were able to further discriminate between real data sets (Section \ref{sec:real1}) where the current methodology could not (Table \ref{tab:BRASS}). The proposed hypothesis tests can also be used to characterise the different types of behaviour present in the data. In the applications provided, we illustrated the important implications in further understanding the mechanisms behind the plant and nematode circadian clocks, and the environmental implications associated with soil pollution. However, we note that our methodology can readily be applied to other circadian datasets, as well as to data originating in other fields.

In all of our proposed hypothesis tests, we wish to test many hypotheses of the type $H_0: S_j^{(1)} (k/T) = S_j^{(2)} (k/T)$ for several values of $j$ and $k$. In this manuscript we adopted the Bonferroni correction and, for a less conservative approach, the false discovery rate (FDR) procedure. Our simulations in Section \ref{sec:sims} showed that both these methods work well. However, the multiple-hypothesis testing methods we use do not account for the dependence of the spectral coefficients. The hypothesis tests developed in Sections \ref{sec:hft} and \ref{sec:ht} alleviate this problem by transforming the data to produce coefficients that are approximately uncorrelated. However, neither method fully decorrelates the data and multiple-hypothesis testing methods that take the dependence of the (transformed) spectral coefficients into account are an interesting avenue of further work.

Finally, we note the wavelet system gives a representation for nonstationary time series under which we estimate the wavelet spectrum and subsequently perform hypothesis testing. Ideally, we would envisage the use of the wavelet that is best suited to modelling and discriminating between the particular dataset. In simulations we found our method to be fairly robust to the wavelet choice. An area of further work would be to derive a procedure for determining which wavelet system to adopt for any given dataset.



\appendix

\section{Experimental Details}
\label{App:Expermiental Methods}
\beginsupplement
In this section we outline the experimental details that led to the datasets introduced in Section \ref{sec:intro} and subsequently analysed in Sections \ref{subsec:app case 1 metals}, \ref{subsec:app case 2 AM} and \ref{subsec:app case 3 worms}.

\noindent{\em Experimental overview: Lead and Ultradian Datasets.} Both Davis and Millar labs used a firefly luciferase reporter system. This involves fusing the gene of interest (here, `cold and circadian regulated and RNA binding 2', \textit{CCR2}) to a bioluminescent enzyme called luciferase \citep{doyle2002elf4}. When \textit{CCR2} is expressed, the resultant luciferase emits light which is measured using a TopCount NXT scintillation counter (Perkin Elmer), allowing relative gene expression of \textit{CCR2} to be quantified \textit{in vivo} \citep{southern2005circadian,perea2015modulation}.

\noindent\textbf{Lead nitrate dataset.}
\textit{Arabidopsis thaliana} seeds (Ws--\textit{CCR2:LUC} \citep{doyle2002elf4}) were surface sterilised and plated onto Hoagland's media containing 1\% sucrose, 1.5\% phyto agar \citep{hoagland1950water}. The seeds were stratified for 2 days at 4$^{\circ}$C and transferred to growth chambers to entrain under 12:12 light/dark cycles at a constant temperature of 20$^{\circ}$C. Six-day-old seedlings were transferred to 96 well microtiter plates containing Hoagland's $1\%$ sucrose, $1.5\%$ agar \citep{hanano2006multiple} with or without supplemental Pb(NO$_3$)$_2$ (lead nitrate) at a concentration of 1.4mM. After 24 hours, the plants were then transferred to the TOPCount machine. Measurements were taken at intervals of approximately 45 minutes. Measurement began after the transition to 12 hours of darkness
(known as subjective dusk) on the seventh day of the plants' life. Therefore, the plants experience one `normal' day in the TOPCount machine (known as entrainment). After this, the plants are exposed to constant light (known as an LL free-run) for approximately four days. This dataset consists of 48 plant signals recorded at $T=128$ time points, with both the `Control' and `Lead' groups containing 24 plants.

\noindent \textbf{Ultradian dataset.} \citep{millar2015changing}.
This dataset was obtained following a similar method as outlined for the Lead dataset above, but compared `Control' \textit{A. thaliana} plants (Ws--2 with \textit{CCR2:LUC} \citep{doyle2002elf4}) with `Mutant' \textit{A. thaliana} plants (Ws--2 \textit{cca1 lhy}). Plants were grown on MS media \cite{murashige1962revised} with 3\% sucrose and 1.5\% phyto-agar. Plants were entrained in 12:12 L:D conditions at 22$^{\circ}$C followed by an LL free-run. Measurements were taken at intervals of approximately 30 minutes. This dataset consists of 48 plant signals recorded at $T=256$ time points, with both the `Control' and `Mutant' groups containing 24 plants.

\noindent \textbf{Nematode dataset.}
This dataset was obtained using male \textit{Caenorhabditis elegans} strain PE254 (obtained from the CGC), which expresses firefly luciferase under the promoter of the \textit{sur-5} gene (\textit{feIs4 [Psur-5::luc+::gfp; rol-6(su1006)]} \cite{lagido2008bridging}). Nematodes expressing luciferase driven by the \textit{sur-5} promoter have previously been reported to show circadian rhythms in luminescence \citep{goya2016circadian}. Single nematodes were placed in wells containing 100$\mu$l S buffer \citep{stiernagle1999maintenance}, supplemented with 5 mg/mL
cholesterol, 1 g/L wet weight pelleted \textit{Escherichia coli} OP50 strain and 100 µM luciferin. Treatment wells also contained 10 $\mu$M SB 203580 (a p38 MAPK inhibitor (Sigma S8307)). Entrainment conditions were 12 hours at 20$^{\circ}$C followed by 12 hours at 15$^{\circ}$C for two days in constant darkness. Free-running was at 20$^{\circ}$C in constant darkness. Luciferase measurements were recorded approximately every 13 minutes. Nematodes that died (shown by a sudden loss of luciferase expression) were excluded from data analysis. Therefore, this dataset consists of 62 signals recorded at $T=512$ time points, with the `Control' and `Treatment' groups containing 32 and 30 time series respectively.

\section{Real data analysis: Supplementary Material}
\label{App:RDASupMat}
In this section, for each motivating example dataset, we report: a summary of the output of the analysis of the motivating datasets in BRASS (Table \ref{tab:BRASS}); the results of the Priestley-Subba Rao test of stationarity (for each time series) in Table \ref{tab:stattest} and the number of rejections for the relevant proposed hypothesis testing  procedure (Table \ref{tab:ResultsRDA}).

\begin{table}
\begin{center}
\begin{tabu}{|c|c|c|c|c|}
\hline \thead{Dataset} & \thead{Mean Period Estimate: \\ Control Group} & \thead{Mean Period Estimate: \\ Test Group} & \thead{Difference} & \thead{p--value} \\
\hline \textbf{Lead}  & 27.4 & 26.8 & -0.6 & 0.16 \\
\hline \textbf{Ultradian}  & 6.5 & 6.5 & 0.0 & 0.98 \\
\hline \textbf{Nematode}  & 24.8 & 25.6 & +0.8 & 0.55 \\
\hline
\end{tabu}
\caption{A summary of the output of the analysis of the motivating example datasets in BRASS: the mean period estimate for the control and test groups in hours (obtained using FFT-NLLS analysis \citep{plautz1997quantitative}), the difference between the period estimates and the corresponding p--value.}
  \label{tab:BRASS}
\end{center}
\end{table}

\begin{table}
\centering
\begin{tabular}{|c|c|c|c|}
\hline \thead{Dataset} & \thead{Lead} & \thead{Ultradian} & \thead{Nematode} \\
\hline  Number of nonstationary time series & 39 ($81\%$) & 41 ($85\%$) &  61 ($98\%$) \\
\hline Total number of time series & 48 & 48 & 62  \\
\hline
\end{tabular}
\caption{Results for the Priestley-Subba Rao test of stationarity, implemented in the
{\tt fractal} package in R and available from the CRAN package repository. Number of nonstationary plants indicates the number of time series (in each motivating example dataset) with enough evidence to reject the null hypothesis of stationarity at the 5\% significance level (as a percentage in brackets).}
\label{tab:stattest}
\end{table}


\begin{table}
\begin{center}
\begin{tabu}{|c|[2pt]c|c|}
\hline \thead{Dataset (Test)} & \thead{Bon.} & \thead{FDR}\\
\hline \textbf{Lead (FT)} & 31 (3\%) & 133 (15\%) \\
\hline \textbf{Ultradian (HFT)} & 1102 (54\%) & 1538 (75\%) \\
\hline \textbf{Nematode (HT)} & 0 (0\%) & 0 (0\%) \\
\hline
\end{tabu}
\caption{The number of rejections (as a percentage in brackets) for each relevant proposed test and multiple-hypothesis testing procedure for the motivating example datasets.}
  \label{tab:ResultsRDA}
\end{center}
\end{table}


\section{Detailed Description of Simulation Studies}
\label{App:SimStudyDetails}
In this section we give a more detailed description of the simulation studies outlined in Section \ref{sec:sims}. The basic structure of each simulated experiment can be described as follows. In each case, we assumed that the signal was a realisation of length $T=256$ from one of $i=1,2$ possible groups, each having (possibly) different spectral structure. A set of $N_1 = N_2 = 1, 10, 25, 50$ signal realisations for each group was generated either from variously defined: spectra (models \textbf{P1}--\textbf{P5} and \textbf{M1} and \textbf{M2}); AR processes (models \textbf{P6}, \textbf{P7}, \textbf{M3} and \textbf{M4}) or functional time series (models \textbf{P8}--\textbf{P12} and \textbf{M5}).

For the models defined by group spectra, signal realisations were generated using the \verb|locits| package in R (available from the CRAN package repository) and the representation in equation \eqref{LSW rep} with the Haar wavelet and a Gaussian orthonormal increment sequence with mean zero and unit variance. (Note that the \verb|wavethresh| package in R preceded the \verb|locits| package and can also be used to generate LSW processes. For more information on how to generate LSW processes from a particular spectrum see \cite{nason2008wavelet}.)

\subsection{Model Details}
In this section we give a detailed description of each model outlined in Sections \ref{subsec:sim power comparison} and \ref{subsec:sim size comparison}.

\begin{enumerate}
 \item \textbf{P1: Fixed Spectra.}
 We follow \cite{krzemieniewska2014classification} Section 4.1.1- Fixed spectra where the spectra of the two groups differ only at the finest level by $100$ coefficients. We simulate each replicate $r_i$-th time series of length $T=256$ of the $i$-th group from the wavelet spectrum $\{ S_{j}^{(i)}(z) \}_{j=1}^{J}$ which we define for each of the $i=1,2$ groups as follows:
 \begin{equation}
 \label{eq:KrezSim1}
    S_j^{(1)}(z)=
     \begin{cases}
       4\cos^2(2 \pi z), & \text{for}\ j = 3, z \in (0, 1)\\
       1, & \text{for}\ j = 7, z \in (1/256, 56/256) \\
       0, & \text{otherwise;}
     \end{cases}
 \end{equation}
 and
 \begin{equation}
 \label{eq:KrezSim2}
    S_j^{(2)}(z)=
     \begin{cases}
       4\cos^2(2 \pi z), & \text{for}\ j = 3, z \in (0, 1)\\
       1, & \text{for}\ j = 7, z \in (1/256, 156/256) \\
       0, & \text{otherwise.}
     \end{cases}
 \end{equation}
Figure \ref{fig:KrezSimpdf} provides a visualisation of the wavelet spectra (top row) and an example of a signal realisation from each of the two groups (bottom row).

 \begin{figure}
 \centering
 \includegraphics[width=\linewidth]{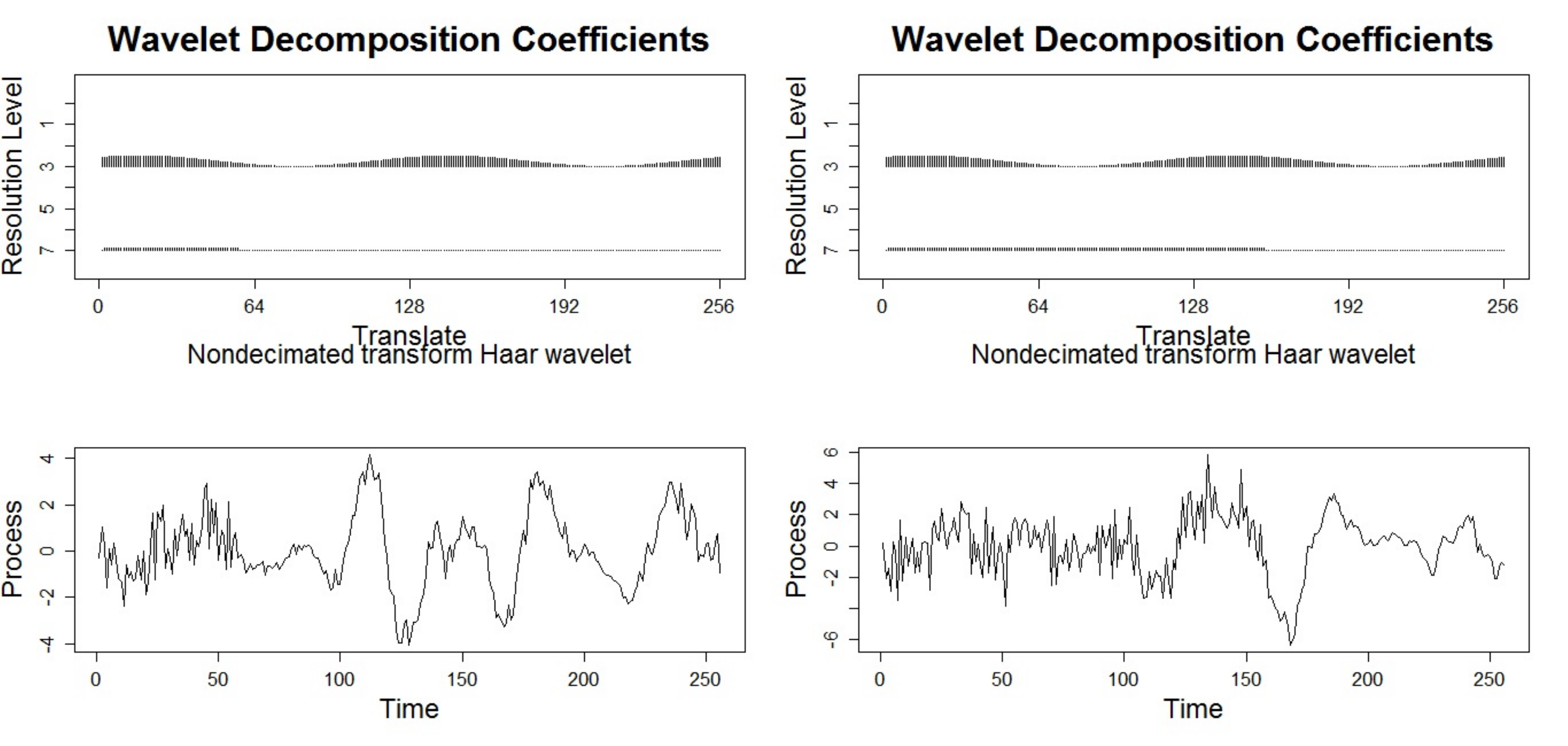}
 \caption{\textbf{P1:Fixed Spectra.} Top left: Group 1 wavelet spectrum; Top right: Group 2 wavelet spectrum; Bottom left: Group 1 realisation; Bottom right: Group 2 realisation.}
 \label{fig:KrezSimpdf}
 \end{figure}

 \item\textbf{P2: Fixed Spectra-Fine Difference.}
 For our next study, we modify the setting above such that the spectra of the two groups differ by $6$ coefficients (in resolution level 7). Therefore, $\{ S_{j}^{(1)}(z) \}_{j=1}^{J}$ is as defined in equation \eqref{eq:KrezSim1} above but we specify the evolutionary wavelet spectrum $\{ S_{j}^{(2)}(z) \}_{j=1}^{J}$ as follows:
 \begin{equation}
 \label{eq:KrezSim2C}
    S_j^{(2)}(z)=
     \begin{cases}
       4\cos^2(2 \pi z), & \text{for}\ j = 3, z \in (0, 1)\\
       1, & \text{for}\ j = 7, z \in (1/256, 50/256) \\
       0, & \text{otherwise.}
     \end{cases}
 \end{equation}

 Figure \ref{fig:KrezSimCpdf} provides a visualisation of the wavelet spectra (top row) and an example of a signal realisation from each of the two groups (bottom row).

   \begin{figure}
   \centering
   \includegraphics[width=\linewidth]{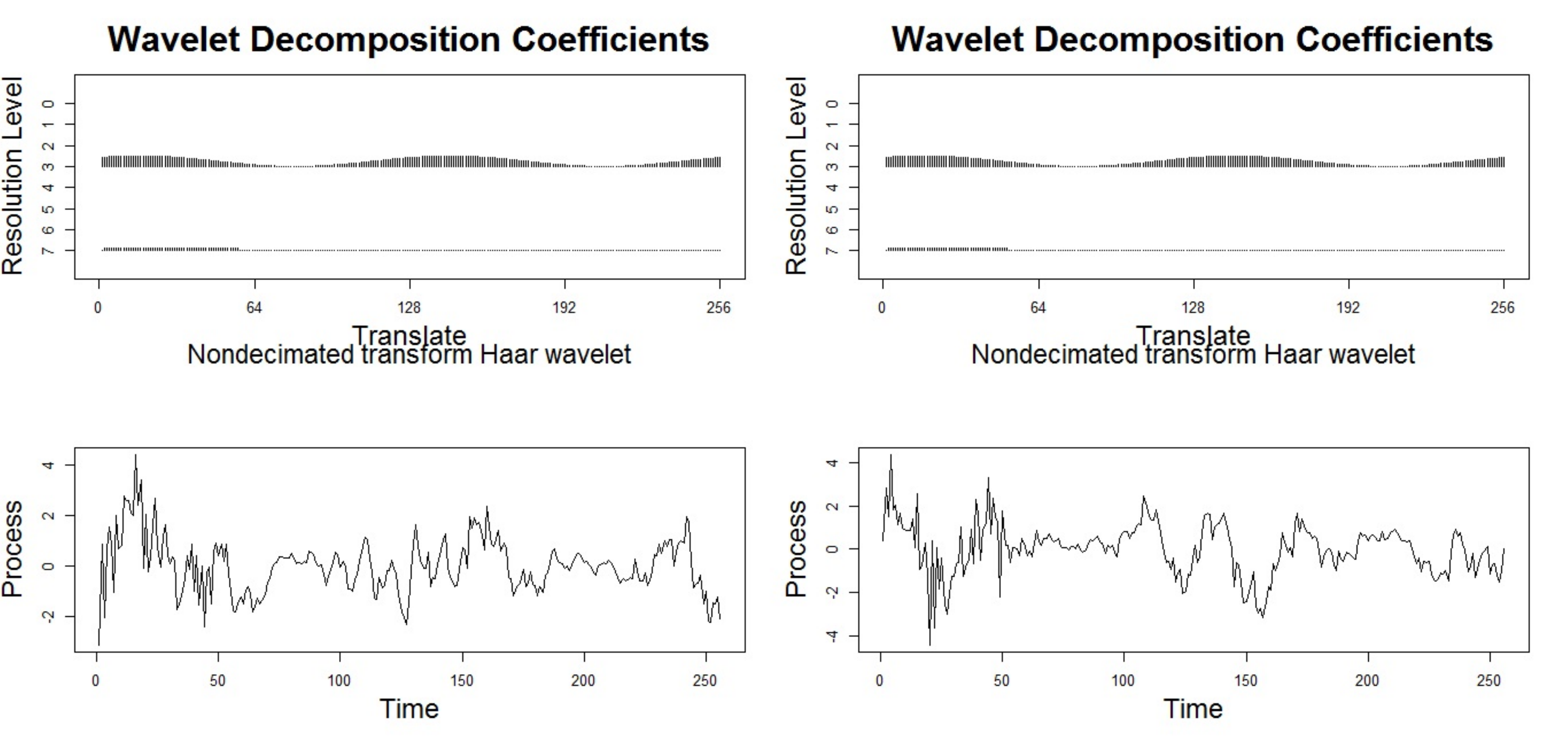}
   \caption{\textbf{P2:Fixed Spectra-Fine Difference.} Top left: Group 1 wavelet spectrum; Top right: Group 2 wavelet spectrum; Bottom left: Group 1 realisation; Bottom right: Group 2 realisation.}
   \label{fig:KrezSimCpdf}
   \end{figure}

 \item \textbf{P3: Fixed Spectra-Plus Constant.}
 We now define fixed spectra such that the spectra of the two groups differ by a constant at the finest resolution level. Therefore, $\{ S_{j}^{(1)}(z) \}_{j=1}^{J}$ is as defined in equation \eqref{eq:KrezSim1} above but we specify the evolutionary wavelet spectrum $\{ S_{j}^{(2)}(z) \}_{j=1}^{J}$ as follows:
 \begin{equation}
 \label{eq:KrezSim6.2}
    S_j^{(2)}(z)=
     \begin{cases}
       4\cos^2(2 \pi z), & \text{for}\ j = 3, z \in (0, 1)\\
       2, & \text{for}\ j = 7, z \in (1/256, 56/256) \\
       1, & \text{for}\ j = 7, z \in (57/256, 256/256) \\
       0, & \text{otherwise.}
     \end{cases}
 \end{equation}
  Figure \ref{fig:KrezSimGpdf} provides a visualisation of the wavelet spectra (top row) and an example of a signal realisation from each of the two groups.

   \begin{figure}
   \centering
   \includegraphics[width=\linewidth]{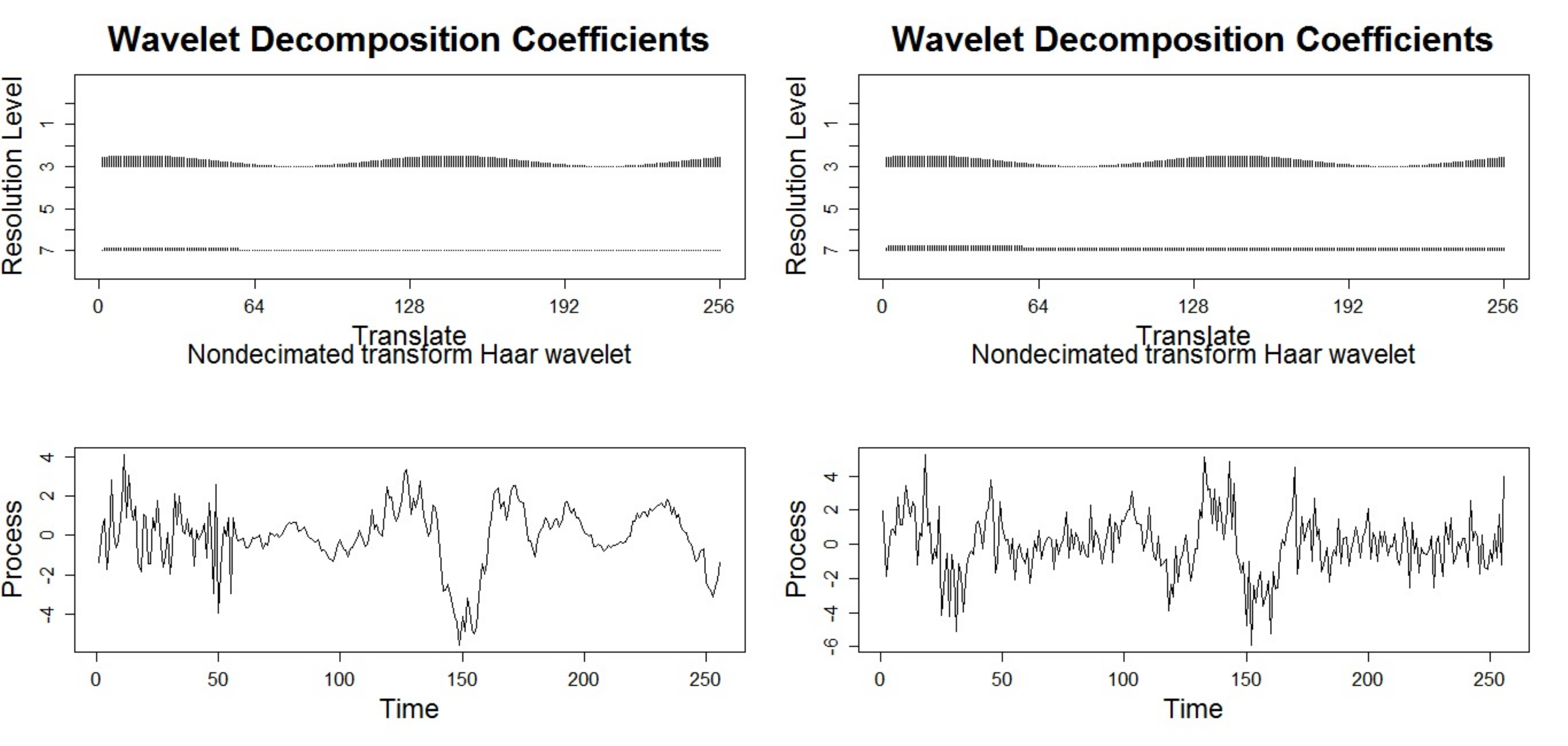}
   \caption{\textbf{P3:Fixed Spectra-Plus Constant.} Top left: Group 1 wavelet spectrum; Top right: Group 2 wavelet spectrum; Bottom left: Group 1 realisation; Bottom right: Group 2 realisation.}
   \label{fig:KrezSimGpdf}
   \end{figure}

\item \textbf{P4/P5: Gradual Period Change.}
With this simulation study aiming to replicate a typical circadian experiment with changes beyond the stationarity assumption, we define time series as realisations from one of $3$ possible groups, each with different spectral characteristics. In particular, each group represents a time series that gradually changes period from 24 to: 25 (Group 1), 26 (Group 2) and 27 (Group 3) over (approximately) two days, before continuing with the relevant period for a further two days. We choose $T=256$ which is equivalent to a free-running period of 4 days with equally spaced observations every $22.5$ minutes. Figure \ref{fig:specs Case 2} shows the wavelet spectra which display the gradually changing periods that define each of the 3 groups. (Note that the increased period is shown by the movement up through the resolution levels and the gradual increase in period of the wavelet coefficients.) To determine which changes can be discriminated by the methods, we perform two studies within this setting: \textbf{P4}: simulations from Group 1 and Group 2 and \textbf{P5}: simulations from Group 1 and Group 3.

\begin{figure}
  \centering
  \includegraphics[width=\linewidth]{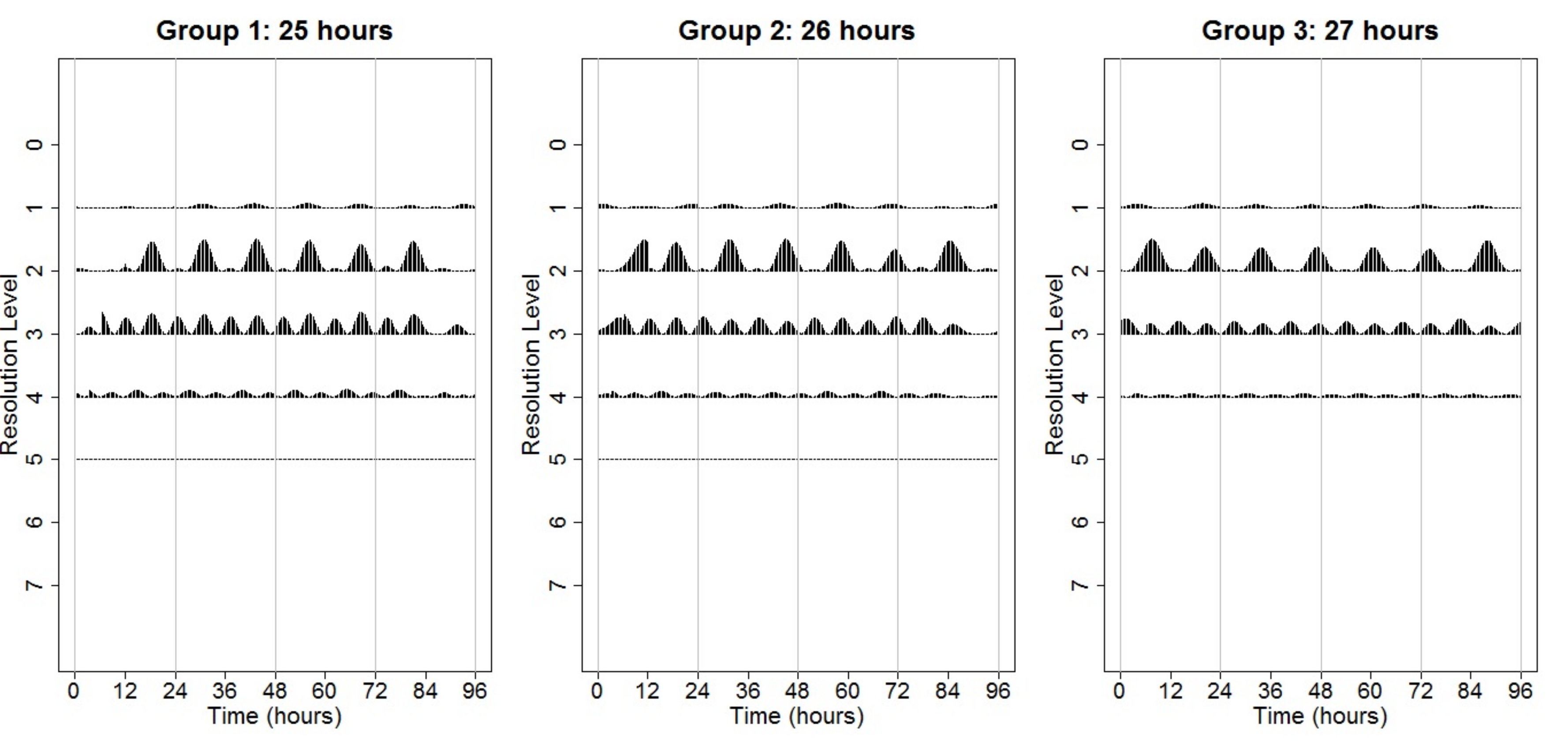}
  \caption{\textbf{P4/P5: Gradual Period Change.} Left: Group 1 wavelet spectrum (gradual period change from 24 to 25 hours); Centre: Group 2 wavelet spectrum (gradual period change from 24 to 26 hours); Right: Group 3 wavelet spectrum (gradual period change from 24 to 27 hours).}
  \label{fig:specs Case 2}
  \end{figure}

\item \textbf{P6/P7: AR Processes with Time-Varying Coefficients.}
The signals in models \textbf{P1--P5} are generated from a defined group spectrum, satisfying the underlying LSW modelling assumptions of our proposed tests. The purpose of this study is to asses the performance of our tests when these assumptions are not met. Therefore, we simulate from an important class of nonstationary processes-- AR processes with time-varying coefficients. We propose a simulation study in a setting as described in \cite{fryzlewicz2009consistent} Section 4.1 Cases 1 and 2.

\textbf{P6: AR Processes with Abruptly Changing Parameters.} The $r_i$-th time series from group $i = 1, 2$, denoted $X_{n,t}^{(i), r_i}$ is generated from the process defined by:
\begin{equation}
\label{eq:Case5.1}
X_{t}^{(i), r_i} = \phi_1^{(i)}(t) X_{t-1}^{(i), r_i} + \phi_2^{(i)}(t) X_{t-2}^{(i), r_i} + \epsilon_{t}^{(i), r_i},
\end{equation}
where the innovations $\epsilon_{t}^{(i), r_i}$ are independent and identically distributed (iid) Gaussian with zero mean and unit variance. In this study, the squared difference between the group spectra is relatively small and the abruptly changing parameters for the two groups are shown in Table \ref{tab:5.1Bparams}. Representative time series plots from each group and the estimated spectra are shown in Figure \ref{fig:FOCMSimpdf}.

\begin{table}
\begin{center}

\begin{tabular}{|c|c|c|c|}
\hline Time-varying parameters & Time Index & Group $i=1$ & Group $i=2$  \\
\hline $\phi_1^{(i)}(t)$ & $t = 1, \dots, 53$ & 0.8 & 0.8  \\
\hline  & $t = 54, \dots, 128$ & -0.9 & -0.3  \\
\hline   & $t = 129, \dots, 256$ & 0.8 & 0.8 \\
\hline  $\phi_2^{(i)}(t)$ & $t = 1, \dots, 256$ & -0.81 & -0.81 \\
\hline
\end{tabular}
\caption{\textbf{P6: AR Processes with Abruptly Changing Parameters.} The abruptly changing parameters of two nonstationary autoregressive processes.}
  \label{tab:5.1Bparams}
\end{center}
\end{table}

 \begin{figure}
  \centering
  \includegraphics[width=\linewidth]{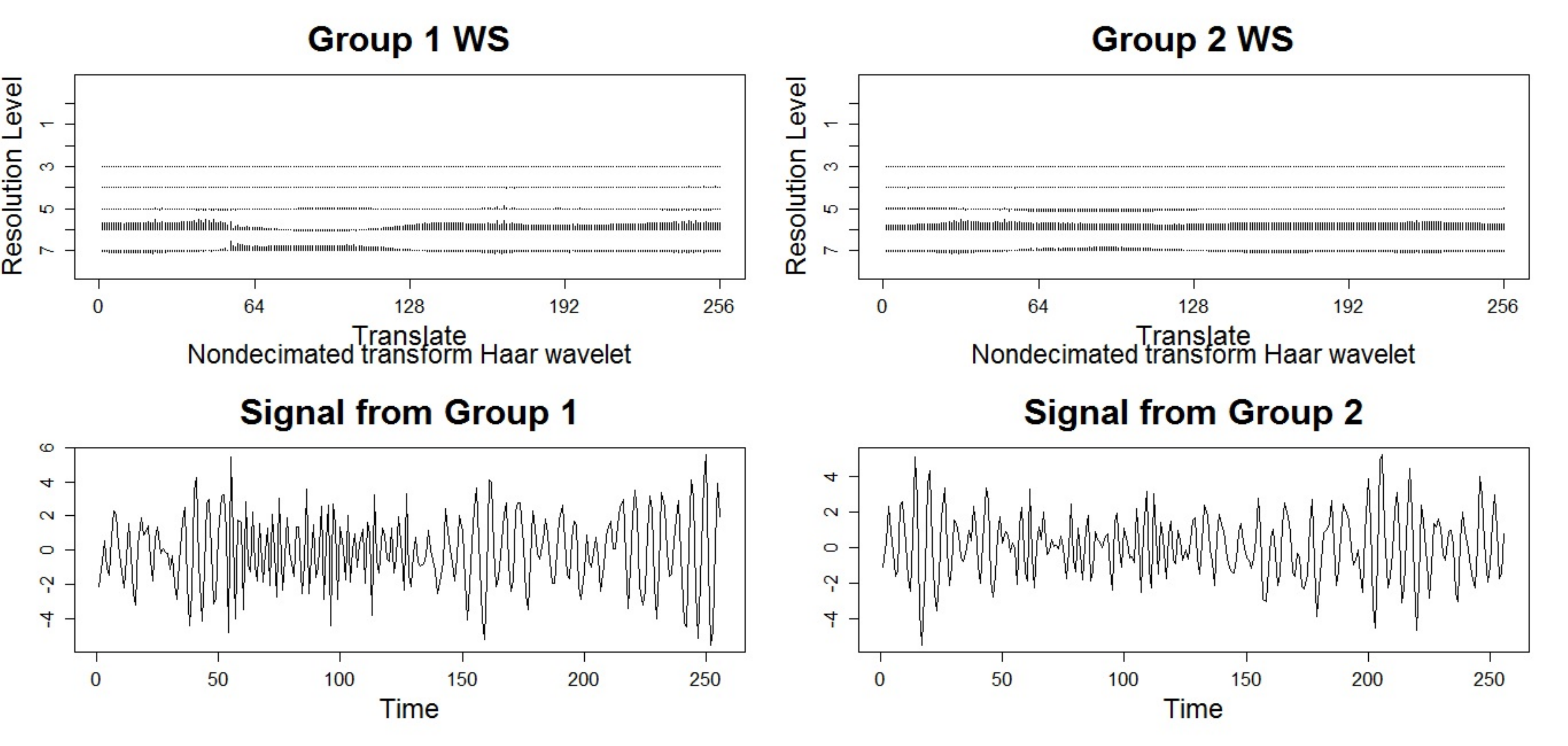}
  \caption{\textbf{P6: AR Processes with Abruptly Changing Parameters.} Nonstationary autoregressive processes. Top left: Estimated wavelet spectrum of Group 1; Top right: Estimated wavelet spectrum of Group 2; Bottom left: Group 1 realisation; Bottom right: Group 2 realisation.}
  \label{fig:FOCMSimpdf}
  \end{figure}

\textbf{P7: AR Processes With Slowly Changing Parameters.} The $r_i$-th time series from group $i = 1, 2$, denoted $X_{t}^{(i), r_i}$ is generated from the process defined by:
\begin{equation}
\label{eq:Case5.2}
X_{t}^{(i), r_i} = \phi_1^{(i)}(t) X_{t-1}^{(i), r_i} + \phi_2^{(i)}(t) X_{t-2}^{(i), r_i} + \epsilon_{t}^{(i), r_i},
\end{equation}
where the innovations $\epsilon_{t}^{(i), r_i}$ are iid Gaussian with zero mean and unit variance. In this study, the group wavelet spectra are highly similar and hence the squared difference between group spectra is relatively small. The slowly changing parameters for groups $i=1,2$ are shown in Table \ref{tab:5.2Bparams}. Representative time series plots from each group and the estimated spectra are shown in Figure \ref{fig:FOCMSim5.2Bpdf}.

\begin{table}
\begin{center}
\begin{tabular}{|c|c|c|}
\hline Time-varying parameters & Group $i=1$ & Group $i=2$  \\
\hline $\phi_1^{(i)}(t)$ &  $-0.8[1-0.7\cos(\pi t/T)]$ & $-0.8[1-0.1\cos(\pi t/T)]$  \\
\hline  $\phi_2^{(i)}(t)$ &  -0.81 & -0.81 \\
\hline
\end{tabular}
\caption{\textbf{P7: AR Processes With Slowly Changing Parameters.} The slowly changing parameters of two nonstationary autoregressive processes.}
  \label{tab:5.2Bparams}
\end{center}
\end{table}

   \begin{figure}
   \centering
   \includegraphics[width=\linewidth]{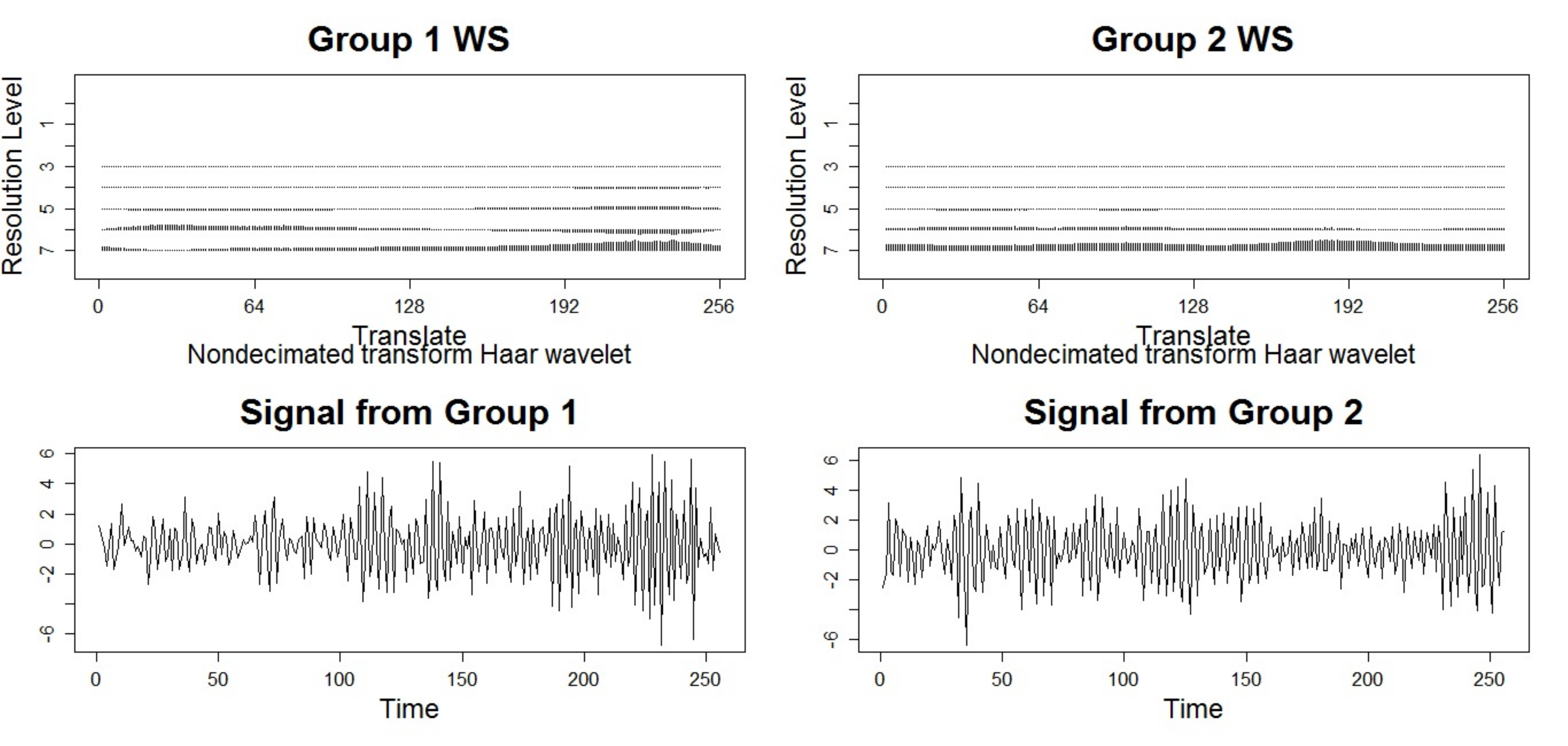}
   \caption{\textbf{P7: AR Processes with Slowly Changing Parameters.} Top left: Estimated wavelet spectrum of Group 1; Top right: Estimated wavelet spectrum of Group 2; Bottom left: Group 1 realisation; Bottom right: Group 2 realisation.}
   \label{fig:FOCMSim5.2Bpdf}
   \end{figure}

\item \textbf{P8--P12: Functional Time Series (Constant Period).}
This study follows \cite{zielinski2014strengths} and generates each time series using an underlying cosine curve with additive noise, which also coincides with the theoretical assumptions of the ANT. As in Models \textbf{P4} and \textbf{P5}, we choose $T=256$, which is equivalent to a free-running period of 4 days with equally spaced observations every $22.5$ minutes.  The $r_i$-th time series from group $i = 1, 2$, denoted $X_{t}^{(i), r_i}$ is generated from the process defined by:
\begin{equation}
\label{eq:FuncTS}
X_t^{(i), r_i} = f^{(i)}(t) + \epsilon^{(i), r_i}_t,
\end{equation}
where the random variables $\epsilon^{(i), r_i}_t$ are iid Gaussian with zero mean and unit variance and the functions $f^{(i)}(t)$ are defined below.
We define time series as realisations from one of 6 possible groups, each with a different (constant) period. The function $f^{(i)}(t)$ is set as a cosine curve with an amplitude of 2 and a period of: 24 hours (Group 1), 21 hours (Group 2), 22 hours (Group 3), 23 hours (Group 4), 23.5 hours (Group 5) and 23.75 hours (Group 6). Representative time series plots and the estimated spectra for Groups 1 and 4 are shown in Figure \ref{fig:FuncTSP8}. To determine which period changes can be discriminated by the methods, we perform five studies within this setting: simulations from Group 1 and Groups 2--6 (models \textbf{P8--P12} respectively).
   \begin{figure}
   \centering
   \includegraphics[width=\linewidth]{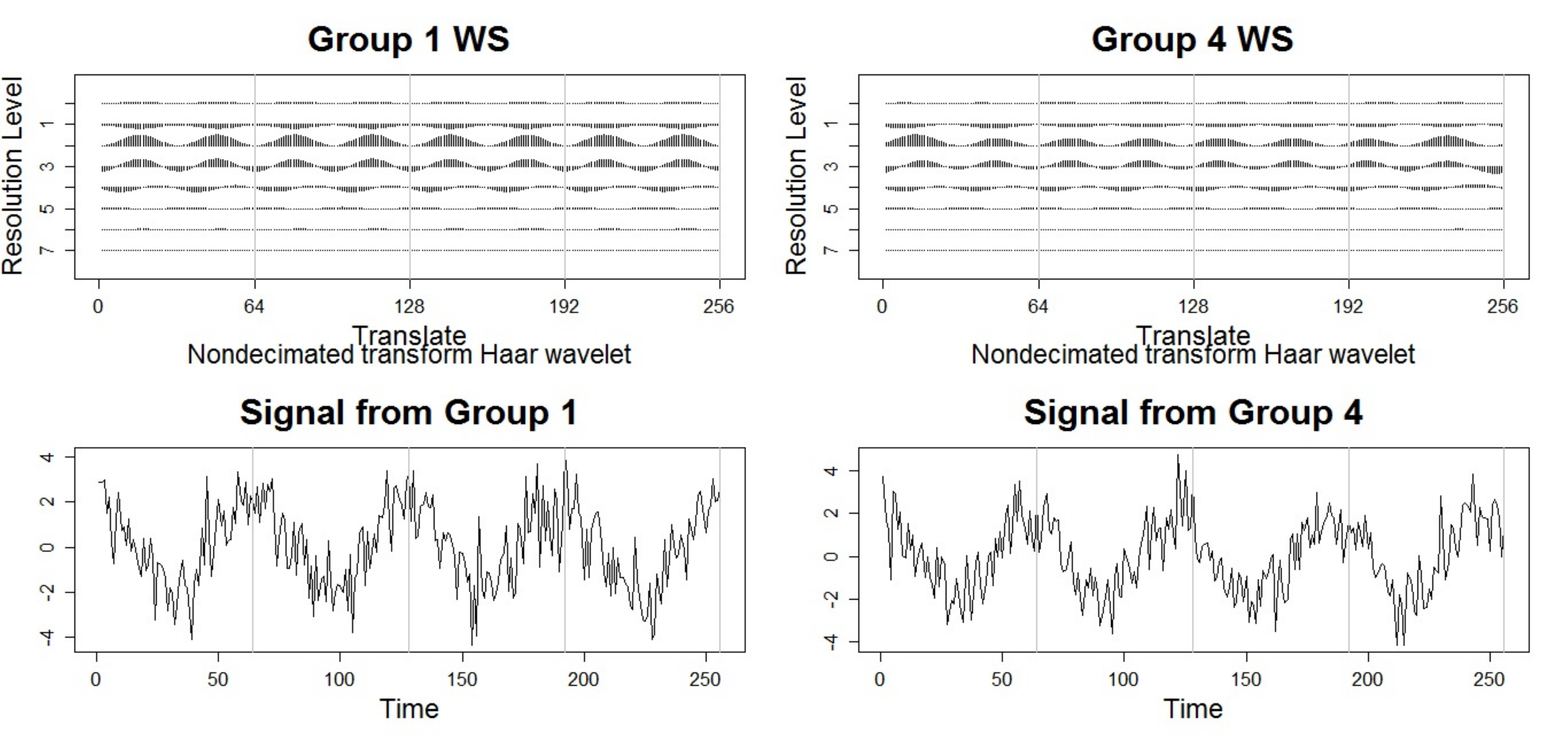}
   \caption{\textbf{P10: Functional Time Series with Constant Period.} Top left: Estimated wavelet spectrum of Group 1 (24 hour period); Top right: Estimated wavelet spectrum of Group 4 (23 hour period); Bottom left: Group 1 realisation; Bottom right: Group 4 realisation. Grey lines indicate a 24 hour period.}
   \label{fig:FuncTSP8}
   \end{figure}
 \end{enumerate}

\subsection{Supplementary Tables}
\label{App:SupTabs}
In this section we provide results which support the discussion of the hypothesis tests in Section \ref{sec:sims}. We report the simulated power and size estimates for $N_1 = N_2 = 1, 10, 50$ for the simulation studies outlined in Sections \ref{subsec:sim power comparison} and \ref{subsec:sim size comparison} in tables \ref{tab:simPower1} -- \ref{tab:Sizesim50}. Additionally, we report the number of rejections for the FT for model \textbf{M4} with $N_1 = N_2 = 10$ and $25$ and both multiple-hypothesis testing methods in Table \ref{tab:SimsCase 5.2B FT rejs}. We also report the simulated power and size estimates for $N_1 = N_2 = 10$ and $25$ for models \textbf{P8--P12} and \textbf{M5} (outlined in Sections \ref{subsec:sim power comparison} and \ref{subsec:sim size comparison}) in Table \ref{tab:sim_Comp}.


\begin{table}
\begin{center}
\begin{tabu}{|c|c|c|c|c|c|c|c|}
\hline \textbf{Model} & \textbf{P1} & \textbf{P2} & \textbf{P3} & \textbf{P4} & \textbf{P5} & \textbf{P6} & \textbf{P7} \\
\hline \textbf{HFT(Bon.)} & 69.4 & 3.8 & 72.6 & 4.1 & 51.3 & 2.5 & 21.8\\
\hline \textbf{HFT(FDR)} & 77.7 & 4.9 & 79.0 & 5.4 & 57.9 & 15.2 & 35.9 \\
\hline
\end{tabu}
\caption{Simulated power estimates ($\%$) for the HFT for models P1-P7 with nominal size of $5\%$ with $N_1 = N_2 = 1$ realisations from each group.}
  \label{tab:simPower1}
\end{center}
\end{table}


\begin{table}
\begin{center}
\begin{tabu}{|c|[2pt]c|c|[2pt]c|c|[2pt]c|c|[2pt]c|c|}
\hline \thead{Model} & \thead{WST \\ (Bon.)} & \thead{WST \\ (FDR)} & \thead{FT \\ (Bon.)} & \thead{FT \\ (FDR)} & \thead{HFT \\ (Bon.)} & \thead{HFT \\ (FDR)} & \thead{HT \\ (Bon.)} & \thead{HT \\ (FDR)} \\
\hline \textbf{P1} & 100.0 & 100.0 & 100.0 & 100.0 & 100.0 & 100.0 & 100.0 & 100.0 \\
\hline \textbf{P2} & 3.5 & 4.6 & \textbf{51.9} & \textbf{54.3} & 4.1 & 6.5 & 16.9 & 17.4 \\
\hline \textbf{P3} & 100.0 & 100.0 & 100.0 & 100.0 & 100.0 & 100.0 & 4.2 & 4.3 \\
\hline \textbf{P4} & 0.5 & 0.6 & 8.4 & 10.8 & 4.8 & 7.0 & \textbf{50.4} & \textbf{55.4} \\
\hline \textbf{P5} & 0.4 & 1.1 & 22.6 & 31.0 & 73.4 & 80.2 & \textbf{95.8} & \textbf{98.4} \\
\hline \textbf{P6} & \textbf{92.2} & \textbf{99.7} & 14.7 & 16.4 & 3.4 & 30.7 & 11.6 & 12.2 \\
\hline \textbf{P7} & \textbf{99.2} & \textbf{100.0} & 11.5 & 12.1 & 30.0 & 54.7 & 75.6 & 77.4 \\
\hline
\end{tabu}
\caption{Simulated power estimates ($\%$) for models P1-P7 with nominal size of $5\%$ with $N_1 = N_2 = 10$ realisations from each group. Highest empirical power estimates are highlighted in bold.}
  \label{tab:sim10}
\end{center}
\end{table}


\begin{table}
\begin{center}
\begin{tabu}{|c|[2pt]c|c|[2pt]c|c|[2pt]c|c|[2pt]c|c|}
\hline \thead{Model} & \thead{WST \\ (Bon.)} & \thead{WST \\ (FDR)} & \thead{FT \\ (Bon.)} & \thead{FT \\ (FDR)} & \thead{HFT \\ (Bon.)} & \thead{HFT \\ (FDR)} & \thead{HT \\ (Bon.)} & \thead{HT \\ (FDR)} \\
\hline \textbf{P1} & 100.0 & 100.0 & 100.0 & 100.0 & 100.0 & 100.0 & 100.0 & 100.0 \\
\hline \textbf{P2}& 94.8 & 97.2 & \textbf{100.0} & \textbf{100.0} & 87.1 & 88.5 & \textbf{100.0} & \textbf{100.0} \\
\hline \textbf{P3} & 100.0 & 100.0 & 100.0 & 100.0 &100.0  & 100.0 & 5.3 & 5.3 \\
\hline \textbf{P4} &  11.8 & 28.0 & 96.0 & 99.0 & 92.0 & 94.8 & \textbf{100.0} & \textbf{100.0} \\
\hline \textbf{P5} & 60.2 & 86.6 & 100.0 & 100.0 & 100.0 & 100.0 & 100.0 & 100.0 \\
\hline \textbf{P6} & 100.0 & 100.0 & 100.0 & 100.0 & 96.7 & 100.0 & 99.3 & 99.8 \\
\hline \textbf{P7} & 100.0 & 100.0 & 99.0 & 100.0 & 100.0 & 100.0 & 100.0 & 100.0 \\
\hline
\end{tabu}
\caption{Simulated power estimates ($\%$) for models P1-P7 with nominal size of $5\%$ with $N_1 = N_2 = 50$ realisations from each group. Highest empirical power estimates are highlighted in bold.}
  \label{tab:sim50}
\end{center}
\end{table}


\begin{table}
\begin{center}
\begin{tabu}{|c|c|c|c|c|c|c|c|}
\hline \textbf{Model} & \textbf{M1} & \textbf{M2} & \textbf{M3} & \textbf{M4} \\
\hline \textbf{HFT(Bon.)} & 2.8 & 4.1 & 0.8 & 1.5 \\
\hline \textbf{HFT(FDR)} & 3.2 & 4.8 & 1.7 & 2.1 \\
\hline
\end{tabu}
\caption{Simulated size estimates ($\%$) for the HFT for models M1--M4 with nominal size of $5\%$ with $N_1 = N_2 = 1$ realisations from each group.}
  \label{tab:simSize1}
\end{center}
\end{table}

\begin{table}
\begin{center}
\begin{tabu}{|c|[2pt]c|c|[2pt]c|c|[2pt]c|c|[2pt]c|c|}
\hline \thead{Model} & \thead{WST \\ (Bon.)} & \thead{WST \\ (FDR)} & \thead{FT \\ (Bon.)} & \thead{FT \\ (FDR)} & \thead{HFT \\ (Bon.)} & \thead{HFT \\ (FDR)} & \thead{HT \\ (Bon.)} & \thead{HT \\ (FDR)} \\
\hline \textbf{M1} & 0.3 & 0.5 & 2.6 & 3.3 & 1.0 & 2.6 & 2.5 & 2.7 \\
\hline \textbf{M2} & 0.0 & 0.2 & 2.4 & 3.6 & 2.0 & 5.0 & 3.3 & 3.3 \\
\hline \textbf{M3} & 0.3 & 1.2 & 4.1 & 4.4 & 0.2 & 1.4 & 1.9 & 2.1 \\
\hline \textbf{M4} & 0.4 & 1.6 & \textbf{5.1} & \textbf{5.6} & 0.9 & 1.8 & 2.1 & 2.2 \\
\hline
\end{tabu}
\caption{Simulated size estimates ($\%$) for models M1-M4 with nominal size of $5\%$ and $N_1 = N_2 = 10$ realisations from each group. Empirical size estimates over the nominal size of $5\%$ are highlighted in bold.}
  \label{tab:Sizesim10}
\end{center}
\end{table}


\begin{table}
\begin{center}
\begin{tabu}{|c|[2pt]c|c|[2pt]c|c|[2pt]c|c|[2pt]c|c|}
\hline \thead{Model} & \thead{WST \\ (Bon.)} & \thead{WST \\ (FDR)} & \thead{FT \\ (Bon.)} & \thead{FT \\ (FDR)} & \thead{HFT \\ (Bon.)} & \thead{HFT \\ (FDR)} & \thead{HT \\ (Bon.)} & \thead{HT \\ (FDR)} \\
\hline \textbf{M1} & 0.4 & 1.1 & 2.4 & 3.9 & 0.3 & 2.4 & 3.1 & 3.3 \\
\hline \textbf{M2} & 0.3 & 0.6 & 3.1 & 3.8 & 1.4 & 3.1 & 2.5 & 2.6 \\
\hline \textbf{M3} & 0.5 & 1.2 & 4.4 & 4.8 & 0.2 & 2.2 & 3.9 & 4.2 \\
\hline \textbf{M4} & 0.2 & 1.1 & 4.4 & 4.8 & 1.3 & 2.6 & 2.8 &  2.9\\
\hline
\end{tabu}
\caption{Simulated size estimates ($\%$) for models M1-M4 with nominal size of $5\%$ and $N_1 = N_2 = 50$ realisations from each group.}
  \label{tab:Sizesim50}
\end{center}
\end{table}


\begin{table}
\begin{center}
\begin{tabu}{|c|[2pt]c|[2pt]c|c|c|c|c|c|[2pt]c|}
\hline \textbf{N} & \thead{Multiple-hypothesis \\ Testing Method} & \textbf{1} & \textbf{2} & \textbf{3} & \textbf{4} & \textbf{6} & \textbf{>10}  & \thead{Modified Empirical \\ Size Estimate} \\
\tabucline[2pt]{-} 10 & Bon. & 44 & 5 & 2 & 0 & 0 & 0 & 0.7 \\
\hline  10 & FDR & 40 & 12 & 3 & 0 & 1 & 0 & 1.6 \\
\tabucline[2pt]{-}  25 & Bon. & 38 & 8 & 0 & 0 & 0 & 0 & 0.8 \\
\hline  25 & FDR & 31 & 16 & 3 & 2 & 0 & 0 & 2.1 \\
\tabucline[2pt]{-}  50 & Bon. & 39 & 5 & 0 & 0 & 0 & 0 & 0.5 \\
\hline  50 & FDR & 32 & 10 & 3 & 0 & 1 & 2 & 1.6 \\
\hline
\end{tabu}
\caption{\textbf{M4: AR Process with Slowly Changing Parameters.} Numbers of rejections in empirical size estimates for the \textbf{Raw Periodogram F-Test} (FT), with Bonferroni Correction (Bon.) and false discovery rate (FDR) and with nominal size of $5\%$. ``Modified Empirical Size Estimate'' is calculated by examining only cases with more than one significant coefficient.}
  \label{tab:SimsCase 5.2B FT rejs}
\end{center}
\end{table}

\begin{table}
\begin{center}
\begin{tabu}{|c|[2pt]c|c|[2pt]c|c|c|c|c|}
\hline \textbf{N} & \textbf{Model}  & \thead{Test Group \\ Period} & \thead{WST \\ (FDR)} & \thead{FT \\ (FDR)} &\thead{HFT \\ (FDR)} & \thead{HT \\ (FDR)} & \thead{ANT}\\
\tabucline[2pt]{-} 10 & \textbf{P8} & 21 & 100.0 & 100.0 & 100.0 & 100.0 & 100.0 \\
\hline 10 & \textbf{P9} & 22 & 100.0 & 100.0 & 93.3 & 100.0 & 100.0 \\
\hline 10 & \textbf{P10} & 23 & 100.0 & 100.0 & 31.9 & 100.0 & 100.0 \\
\hline 10 & \textbf{P11} & 23.5 & 100 & 96.1 & 9.5 & 99.4 & 100.0 \\
\hline 10 & \textbf{P12} & 23.75 & 81.2 & 14.6 & 5.6 & 32.4 & 100.0 \\
\hline 10 & \textbf{M5} & 24 & 2.0 & 2.1 & 3.1 & 4.1 & 7.9 \\
\tabucline[2pt]{-} 25 & \textbf{P8} & 21& 100.0 & 100.0 & 100.0 & 100.0 & 100.0 \\
\hline 25 & \textbf{P9} & 22 & 100.0 & 100.0 & 100.0 & 100.0 & 100.0 \\
\hline 25 & \textbf{P10} & 23 & 100.0 & 100.0 & 92.0 & 100.0 & 100.0 \\
\hline 25 & \textbf{P11} & 23.5 & 100.0 & 100.0 & 31.8 & 100.0 & 100.0 \\
\hline 25 & \textbf{P12} & 23.75 & 100.0 & 97.9 & 9.1 & 98.3 & 100.0 \\
\hline 25 & \textbf{M5} & 24 & 3.0 & 2.7 & 2.7 & 3.5 & 4.8 \\
\hline
\end{tabu}
\caption{Simulated size and power estimates ($\%$) for models P8-P12 and M5 with nominal size of $5\%$ and using the false discovery rate procedure (FDR). Note: Control group period is 24 hours in each model.}
  \label{tab:sim_Comp}
\end{center}
\end{table}

\end{document}